# TWO-DIMENSIONAL KAGOME MATERIALS: THEORETICAL INSIGHTS, EXPERIMENTAL REALIZATIONS, AND ELECTRONIC STRUCTURES

Zhongqin Zhang[1,2,†], Jiaqi Dai[1,2,†], Cong Wang[1,2], Hua Zhu[1,2], Fei Pang[1,2], Zhihai Cheng[1,2] and Wei Ji[1,2,*]

[1]*Beijing Key Laboratory of Optoelectronic Functional Materials & Micro-Nano Devices, School of Physics, Renmin University of China, Beijing 100872, China*

[2]*Key Laboratory of Quantum State Construction and Manipulation (Ministry of Education), Renmin University of China, Beijing 100872, China*

*Emails: wji@ruc.edu.cn (W.J.)

**ABSTRACT:**

In recent years, kagome materials have attracted significant attention due to their rich emergent phenomena arising from the quantum interplay of geometry, topology, spin, and correlations. However, in the search for kagome materials, it has been found that bulk compounds with electronic properties related to the kagome lattice are relatively scarce, primarily due to the hybridization of kagome layers with adjacent layers. Therefore, researchers have shown increasing interest in the discovery and construction of two-dimensional (2D) kagome materials, aiming to achieve clean kagome bands near the Fermi level in monolayer or few-layer systems. Substantial advancements have already been made in this area. In this review, we summarize the current progress in the construction and development of 2D kagome materials. We begin by introducing the geometric and electronic structures of the kagome lattice model and its variants, followed by discussions on the experimental realizations and electronic structure characterizations of 2D kagome materials. Finally, we provide an outlook on the future developments of 2D kagome materials.

**Keywords:** kagome materials, two-dimensional, monolayer, few-layer, kagome bands

# 1. Introduction

A kagome lattice is a crystal structure made up of interlaced triangles and hexagons. Its intriguing electronic properties, including unique band structures and spin frustration, make it an ideal platform for exploring novel phenomena such as electronic correlations, topological effects and quantum magnetism[1–3]. Research on kagome systems has made significant progress, revealing various novel phenomena, including superconductivity[4–6], charge density waves[7–9], and magnetic Weyl semimetals[10–13], which has greatly fueled interest in exploring kagome materials.

However, researchers found that materials with electronic properties related to the kagome lattice are relatively rare. For example, theoretical predictions indicate that out of 3742 known materials with kagome networks, only about 7% exhibit properties related to the kagome lattice[14]. One reason for this difficulty is that kagome lattices are inherently two-dimensional (2D), yet past research has focused primarily on three-dimensional bulk materials. In these bulk materials, the kagome layers are often covered by other layers[4,11,14,15], and interlayer interactions tend to hybridize the electronic states related to kagome layers with those of other layers[4,11], pushing the kagome bands away from the Fermi level or even causing them to eliminate[14,15]. Therefore, in pursuit of neater kagome systems, researchers have shown increasing interest in constructing 2D kagome materials.

A direct approach to constructing 2D kagome lattices involves fabricating monolayer counterparts of kagome materials, such as monolayer $AV_3Sb_5$ (A = Cs, K, Rb) and monolayer $Nb_3X_8$ (X = Cl, Br, I). Theoretical predictions suggest that these monolayers possess novel physical properties[16–18]. However, in practice, both direct exfoliation and molecular beam epitaxy (MBE) growth have encountered technical challenges. Therefore, only a few van der Waals kagome materials, such as $Nb_3SeI_7$[19][20][21] and $Pd_3P_2S_8$[22], have been exfoliated into monolayers, and research on their kagome electronic properties is still lacking.. In contrast, constructing kagome lattices directly within 2D systems, independent of well-studied bulk kagome materials, represents another significant direction. Given the substantial progress in this area, we believe it is time for a comprehensive, interim review.

This review presents the current advancements in the construction and characterization of 2D kagome materials, with a primary focus on experimental progress while also addressing theoretical explanations and some predictions. In section 2, we introduce the structure and band characteristics of the kagome lattice model and its variants. In the following four sections, we sequentially summarize the progress and challenges in constructing 2D kagome lattices in organic systems on surfaces (Section 3), inorganic systems on surfaces (Section 4), moiré systems (Section 5), and superatom systems formed from mirror twin boundaries (Section 6), presented in a chronological order of their development within the field of two-dimensional kagome materials. Additionally, we propose a universal strategy for constructing kagome lattices from triangular lattices in Section 7. In the last section, we outlook on future developments of 2D kagome materials.

## 2. Kagome lattices and theoretical models

The kagome lattice has various structural variants[23–30], some of which can significantly modify the standard band structure. Therefore, this section introduces the structural evolution from the regular kagome lattice to its various variants and how these structural variations affect their band structures.

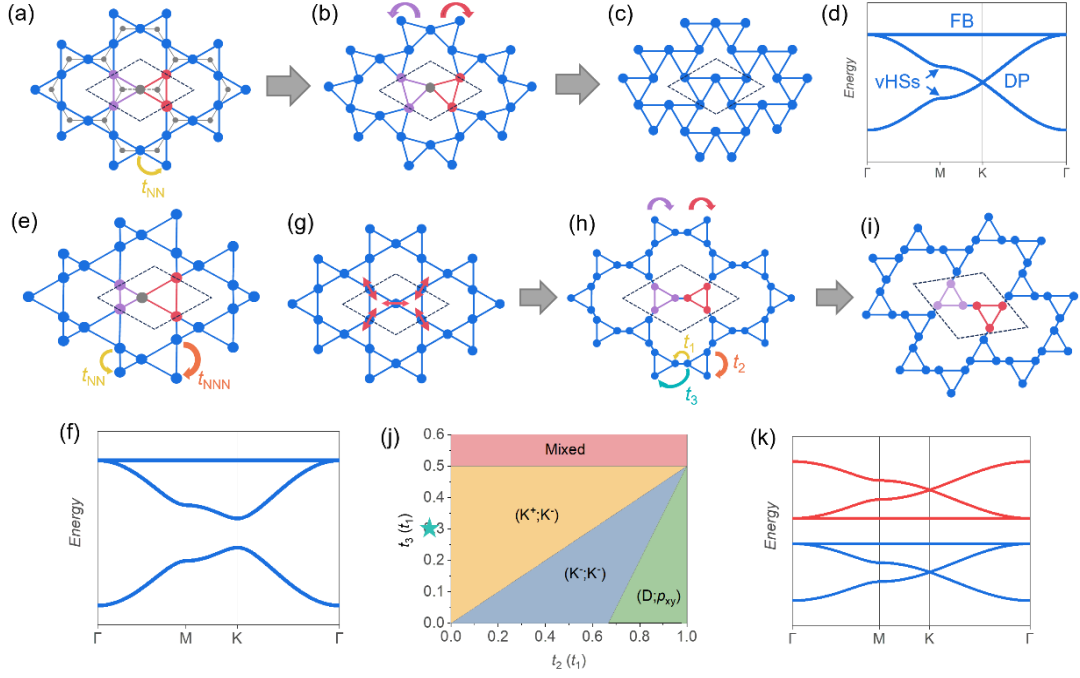

Figure 1 Structure and tight-binding bands of kagome lattices. (a-d) Schematic illustrations of the structures for the regular kagome lattice (a), twisted/distorted kagome lattice (b), and coloring-triangle lattice (c), along with their corresponding band structure (d). In (a), $t_{NN}$ represents the nearest-neighbor hopping constant. In (b), the arrows indicate the rotation direction of the corresponding coloring-triangles in the unit cell. (e-f) Schematic illustrations of the structure (e) and bands (f) of the breathing kagome lattice. In (e), $t_{NN}$ and $t_{NNN}$ represent the nearest-neighbor and next-nearest-neighbor hopping constants, respectively. The band structure shown in (f) corresponds to $t_{NN}/t_{NNN} = 1.5$. (g-k) The structural evolution from the regular kagome lattice (g) to the diatomic kagome lattice (h) and chiral diatomic kagome lattice (i), along with the phase diagram of bands (j) and a band structure [$t_3/t_1 = 0.3$ and $t_2 = 0$, corresponding to the parameters at the star in (j)] (k). In (h), $t_1$, $t_2$, and $t_3$ represent the hopping constants; the arrows indicate the rotation direction of the corresponding coloring-triangles in the unit cell. The dashed line represents the Fermi level when the on-site energy is zero in the tight-binding calculation, and different colors in the band structure indicate two sets of kagome bands. The black dashed lines in the structural schematic diagrams represent the unit cell.

In Figure 1a, the gray lines outline a two-dimensional honeycomb lattice. If the single gray lattice point in the honeycomb lattice is replaced by three blue points arranged in a corner-sharing manner as shown in Figure 1a, the resulting lattice composed of the blue points forms a regular kagome lattice. Each unit cell (marked by the gray dashed diamond) of the regular kagome lattice consists of two corner-sharing triangles, marked in purple and red. The unique structure of the kagome lattice gives it interesting magnetic properties. For example, if the nearest neighbor lattice points are antiferromagnetic coupled, spin frustration occurs. Theoretically, it was predicted that the ground state of the $S = 1/2$ Kagome antiferromagnetic Heisenberg model is a spin liquid[31].

If we only consider the nearest-neighbor hopping ($t_{NN}$) between the kagome lattice points, the tight-binding Hamiltonian in the absence of spin-orbit coupling can be expressed as follows:

$$H(k) = \begin{pmatrix} 0 & 2t\cos k_1 & 2t\cos k_2 \\ 2t\cos k_1 & 0 & 2t\cos k_3 \\ 2t\cos k_2 & 2t\cos k_3 & 0 \end{pmatrix}$$

where $k_n = \boldsymbol{k} \cdot \boldsymbol{a}_n$, $\boldsymbol{k}$ and $\boldsymbol{a}_n$ represent wavevectors in the reciprocal space and the displacement vectors between the nearest neighbors in the kagome lattice. Solving this Hamiltonian yields the band structure shown in Figure 1d, which features three key characteristics: a flat band (FB) spanning the Brillouin zone, a Dirac point (DP) at the K points, and two van Hove singularities (vHSs) at the M points. When these unique electronic features are near the Fermi level, they may give rise to novel properties in the materials. The flat band in the kagome lattice arises from the destructive quantum interference of the wave functions. In this flat band, the suppression of kinetic energy enhances the ratio of Coulomb interactions ($U$) to hopping constant ($t$), namely $U/t$, leading to the decisive role of Coulomb interactions in determining the material's properties. As a result, the flat band in the kagome lattice introduces strong electronic correlations, leading to a wide spectrum of novel physical phenomena. For instance, experiments have shown that the ferromagnetic ground state in $Fe_3Sn_2$ originates from correlation effects enhanced by the flat band[32]. The Dirac fermions impart topological properties to materials[33], such as magnetic Weyl semimetal phase in $Co_3Sn_2S_2$[10–13]. The vHSs are also major sources to introduce strong electronic correlations and lead to Fermi surface nesting, which in turn may cause instabilities in the Fermi surface. Such instabilities have been observed in $CsV_3Sb_5$, where the competition between charge density waves, superconductivity, and other long-range orders has been reported[6,7,34].

Kagome bands exhibit interesting topological properties. When considering spin-orbit coupling effects, the Dirac points and the intersection between the flat band and the Dirac band will open gaps, which are $Z_2$ topologically nontrivial. In addition, the three kagome bands have topological Chern numbers of $C = \pm1, 0, \mp1$ from top to bottom in energetic order. These topologically nontrivial bands and thire resulting electronic correlation effects render the kagome lattice an ideal platform for studying the interplay of electronic correlations, topology, and magnetism.

As shown in Fig. 1b, after the purple and red triangles each rotate oppositely around their centers (as indicated by the colored arrows in Fig. 1b), the resulting structure is termed a twisted kagome lattice (also referred to as the distorted kagome) (Fig. 1b). If the rotation angle is exactly 30° in opposite directions, the newly forming lattice is referred to a coloring-triangle (CT) lattice (Fig. 1c). The CT lattice existed before it was recognized. For example, each iodine (I) atomic layer in $CrI_3$ forms a CT lattice[35]. Theoretically, the Hamiltonians of the twisted kagome lattice and CT lattice are unitary transformable to the regular kagome lattice's, so they share the same band structure (Figure 1d) [29,30]. In a more general case, if the purple and red triangles are inequivalent in terms of size or interactions, the lattice becomes a breathing kagome lattice (Fig. 1e), introducing further structural complexity and changes of bandstructures. This inequivalence causes the two Dirac bands, originally degenerate at the K (Dirac) point, to split, thereby opening a gap for the Dirac cone.

When each lattice point in the regular kagome lattice is split into two points (Figures 1g and 1h), the resulting structure is referred as a diatomic kagome lattice[23,27,28] (Figure 1h). This lattice can also be regarded as the purple and red triangles in the regular kagome lattice being pulled apart, causing their shared corner points evolving into two separate points. By further rotation of the purple and red triangles, structural chirality is thus introduced into diatomic kagome lattices, forming chiral (diatomic) kagome lattices. A representative of these lattices is shown in Figure 1i.

The diatomic kagome lattice has more complex bandstructures, depending on the relative magnitudes of the three near-neighbor hopping constants ($t_1$, $t_2$, and $t_3$). The relative ratio of these constants results in several combinations of two flat-bands and four Dirac bands in band structures, as illustrated in a phase diagram (Figure 1j). Notably, in the (K+, K-) region (shaded in yellow in Fig. 1j), the band structure features two sets of kagome bands, with the flat bands positioned adjacent to each other. When the Fermi level lies between these two flat bands, theoretical calculations predicted that this unique "yin-yang" flat band structure can facilitate the realization of interesting states or phenomena, such as excitonic insulators[36] and the quantum anomalous Hall effect[37]. The introduction of chirality in the chiral diatomic kagome lattice does not change the fundamental band structure of the diatomic kagome lattice[38–40]. This robustness provides a degree of tolerance for material imperfections, allowing the realization of the kagome band properties even in materials that do not have a perfectly ordered diatomic kagome lattice. However, it is important to note that the tight-binding model discussed above considers only a single orbital at each lattice point and identical nearest-neighbor hopping strengths between them. Introduction of more complex symmetries of orbitals, more orbitals, or longer-range hopping interactions may reshape the appearance of these three bands[24,41]. For instance, if each lattice point in the regular kagome lattice is associated with a $p_x$ or $p_y$ orbital, both the flat band and the Dirac point will disappear[24]. Moreover, when additional orbitals are incorporated into the model, the kagome bands would undergo further alterations[41]. Considering that previous theoretical predictions suggested many interesting phenomena can be realized in multi-orbital honeycomb lattices[42–45], such as quantum anomalous hall effect[46], the multi-orbital kagome lattice may be worth further exploration.

These different types of kagome lattices introduced above can also be combined to form new structures, such as the breathing chiral diatomic kagome lattice[38], which exhibits a band structure that combines the characteristics of both the breathing kagome lattice and the diatomic kagome lattice. Currently, the regular kagome lattice, twisted kagome lattice, CT lattice, and breathing kagome lattice have all been realized in two-dimensional atomic lattices. However, the diatomic and chiral diatomic kagome lattices have only been demonstrated in electronic states and have yet to be realized in atomic structures, calling further research and development.

Considering the unique properties of the kagome lattice mentioned above, the following sections focus on constructing the kagome lattice in real few-layer or monolayer materials. This involves the synthesis of kagome materials and the characterization of their electronic structure properties.

# 3. Organic kagome monolayers on surfaces

Organic molecules, due to their diverse structures, flexible assembly, and tunable functionalities, have been playing an increasingly important role in the design and creation of low-dimensional materials. Shapes and sizes of molecules, as well as types and positions of their functional groups can be precisely controlled. These features allow them to serve as building blocks to construct complex surface lattice structures through methods like surface supramolecular self-assembly and surface chemical reactions[47–52]. To date, tens, even hundreds, of monolayer organic lattices have been successfully synthesized on solid surfaces. Depending on the interactions between molecules, these lattices are categorized into hydrogen-bonded organic frameworks (HOFs), metal-organic frameworks (MOFs), and covalent organic frameworks (COFs) and others like halogen-bonded organic frameworks (XOFs). Over the past two decades, the experimental synthesis of organic kagome lattices from bottom-up has made significant progress. Additionally, theoretical calculations have predicted that various intriguing states and phenomena existing in freestanding monolayer organic kagome lattices, such as superconductivity[53], the quantum anomalous Hall effect[54], excitonic insulators[36], topological insulators[55,56], and quantum spin liquids[57]. All these experimental and theoretical advances have established organic kagome monolayers as a potential material platform for exploring novel quantum states.

This section focuses on two primary bottom-up approaches for constructing organic kagome monolayers on surfaces. One is the direct approach, which involves linking organic monomers on solid-liquid or solid-vacuum interfaces to directly form kagome lattice networks in atomic structures using the molecules themselves (referred as atomic kagome lattices). The other is the indirect approach, where organic molecules form repulsive barriers that spatially confine the surface electronic states of metal substrates, leading to the formation of electronic kagome lattices.

## 3.1. Surface-supported organic kagome monolayer networks

The construction process of organic kagome monolayers on solid surfaces typically progresses from solid-liquid to metal-ultra-high-vacuum (UHV)- interfaces, transitioning from weakly bonded self-assembly to strongly bonded structures that require chemical reactions, *i.e.* moving from simpler to more complex methods. This subsection broadly follows this step-by-step progression to discuss the development of organic kagome lattices. A discussion is given at the end of this subsection on the challenges in the synthesis, electronic characterization, and tuning and application of physical properties of organic kagome lattices.

3.1.1. Solid-liquid interfaces

The growth conditions for growing two-dimensional organic crystals at solid-liquid interfaces are less demanding, compared to these for on metal surfaces under ultra-high vacuum as discussed in subsection 3.1.2~3.1.5. As a result, kagome lattices with various bonding types were often first synthesized at solid-liquid interfaces[51,58,59]. This subsection introduces these pioneering works.

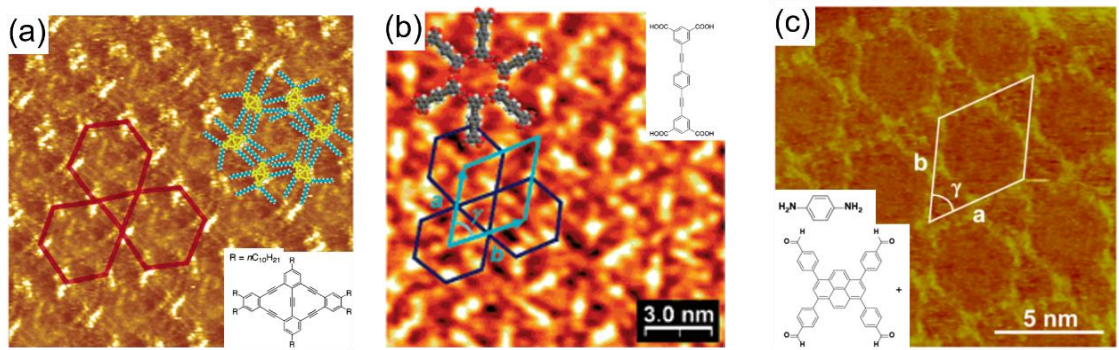

Figure 2 (a) STM image of a kagome network formed by DBA derivatives (inset at bottom right) under ambient conditions. $I_{set}$ = 0.5 nA, $V_{bias}$ = -1.04 V. The scan area is 20.0 × 20.0 nm$^2$. The molecular model of the building block is overlaid on the STM image to aid visualization. The red lines highlight the kagome symmetry. (b) STM image of a kagome lattice constructed from 5,5'-(1,4-Phenylenedi-2,1-ethynediyl)bis(1,3-benzenedicarboxylic acid) molecules (inset at bottom right). $I_{set}$ = 50 pA, $V_{bias}$ = -1.5 V. The unit cell is outlined with blue lines, and a molecular model of the building block is overlaid on the STM image to aid visualization. The black lines highlight the kagome symmetry. (c) STM topography image of a kagome network formed by the condensation of p-phenylenediamine molecules and 1,3,6,8-tetrakis(p-formylphenyl)pyrene molecules (inset) under ambient conditions. $V$ = 700 mV, $I$ = 500 pA. (a) Reproduced from Ref.[51], (b) Reproduced from Ref.[59], (c) Reproduced from Ref.[58].

Van der Waals (vdW) interactions are widespread among organic molecules and often govern their assembly patterns. The first organic kagome network was, to the best of our knowledge, assembled through vdW interactions in 2006. Feyter, Tobe and coworkers used a rhombic derivative of dehydrobenzo[12]annulene (DBA) (inset of Figure 2a) as the building block at a solid-liquid interface formed between highly oriented pyrolytic graphite (HOPG) and 1,2,4-trichlorobenzene (TCB) . Through vdW interactions among them, the molecules self-assembled into the first molecular aggregate in a two-dimensional kagome lattice, which was confirmed by scanning tunneling microscopy (STM) under ambient conditions at the solid-liquid interface[51] (Figure 2a).

One year after, a stronger non-covalent interaction, i.e. hydrogen bonding, was introduced into self-assembly of two-dimensional organic kagome monolayers. In 2007, Wuest and coworkers, employing 5,5'-(1,4-Phenylenedi-2,1-ethynediyl)bis(1,3-benzenedicarboxylic acid) molecules (Fig. 2b inset), synthesized and characterized the first hydrogen-bonded organic framework (HOF) monolayer with a kagome lattice at a HOPG solid-liquid interface[59].

In cases where even stronger intermolecular bonding is involved, such as covalent bonding, the high stability of the bonded molecules often leads to poor crystallinity, making it challenging to form long-range ordered crystals[60]. Despite extensive efforts to synthesize 2D covalent organic kagome lattices[61], it was not until 2017 that the first covalent organic framework (COF) kagome monolayer was successfully synthesized at a solid-liquid interface using dynamic covalent bonds[58]. Wang and coworkers mixed 1,3,6,8-tetrakis(p-formylphenyl)pyrene and p-phenylenediamine (Fig. 2c inset) in a solvent and deposited the mixture onto a HOPG surface. The resulting kagome network, formed through a Schiff base reaction between the two molecules, was imaged using STM under ambient conditions, as shown in Figure 2c.

Although synthesizing two-dimensional kagome lattices at solid-liquid interfaces is relatively straightforward, studying their properties remains challenging. Some surface characterization techniques like STM have reduced resolution at solid-liquid interfaces (as shown in Figure 2) compared to under UHV conditions, and some others are inapplicable in this environment. Thus, the electronic bandstructures of these synthesized organic kagome monolayers are difficult to characterize, which obstruct further investigation on electron- and spin-related phenomena of kagome lattices. To address this issue, the synthesis environment of organic kagome monolayers was expended to metal surfaces in UHV (solid-vacuum interfaces). The following subsections summarize the advances of synthesizing organic kagome lattices on metal surfaces and the studies of their electronic properties in UHV, including self-assembly through vdW forces, hydrogen bonds, and covalent bonds between organic molecules, as well as the introduction of metal atoms to form metal-organic coordination bonds between molecules.

## 3.1.2. VdW-bonding kagome monolayers on metal-vacuum interfaces

Like the case at solid-liquid interfaces, the first organic kagome network at metal-vacuum interfaces was also assembled via vdW interactions. In 2008, Barth, Ruben, Schlickum and coworkers deposited a series of linear dicarbonitrile-polyphenyl (NC-Ph$_5$-CN) molecules (Fig. 3a lower inset) on an Ag(111) surface. Images acquired in UHV STM (Fig. 3a) manifest that the NC-Ph$_5$-CN molecules self-assembled into a kagome lattice through vdW interactions[62], initializing research on organic kagome monolayers on metal surfaces in UHV conditions. Additionally, the triangles forming the kagome lattice, influenced by the polarity of the cyano-groups (CN groups) in NC-Ph$_5$-CN molecules, underwent slight rotations, resulting in a chiral molecular structure (Fig 3a up-right inset), which is a common phenomenon in organic kagome systems[63][64–66]. However, this chirality arises from the structural chirality of the molecules themselves and does not necessarily correspond to the chiral diatomic kagome lattices discussed in Section 2 at the electronic structure level. Subsequently, researchers made a series of outstanding contributions in constructing kagome monolayers using vdW interactions[47,52,67–69], which will not be discussed in detail here.

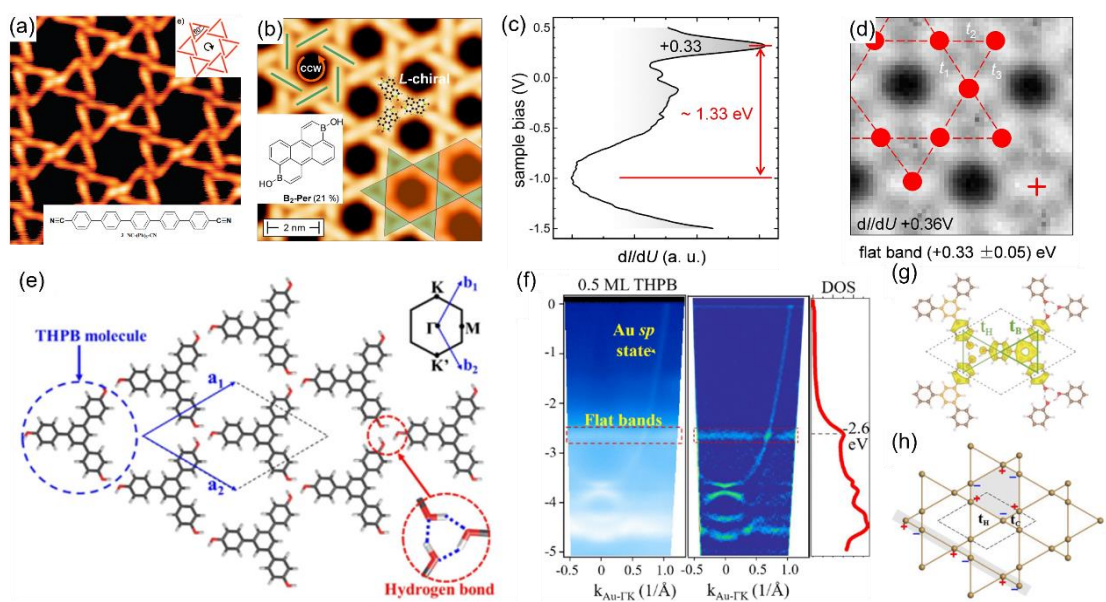

Figure 3 (a) STM image of a kagome network formed by NC-Ph$_5$-CN (inset at bottom right). $I$ = 0.1 nA, $V$ = 0.5 V. (b-d) STM image (b), d$I$/d$V$ spectra (c), and d$I$/d$V$ map at 0.36 V (d) of the kagome lattice constructed by $B_2$-Per. In (b), $U = 1$ V, CCW indicates counterclockwise; the atomic model of building blocks is overlaid on the STM image; green and orange shading represent the triangles and hexagons of the kagome lattice, respectively. In (d), the kagome lattice and lattice points are overlaid on the d$I$/d$V$ map, with the red cross in the bottom right indicating the position where the d$I$/d$V$ spectra were taken. (e-h) (e) Schematic model of the kagome lattice formed by THPB. (f) ARPES spectrum of a 0.5 ML THPB film along the Γ-$K_{Au}$ direction (left), second-derivative plot (middle), and the integrated DOS from the ARPES (right). (g) The optimized lattice structure overlaid with partial charge density derived from the top three valence bands belonging to the breathing-kagome lattice formed by CBRs of THPB. The brown or orange, red, and pink balls represent C, O, and H atoms, respectively. The charge density is plotted using an isosurface of 0.002 eV/Å$^3$. (h) Illustration of an electronic breathing-kagome lattice formed by different hopping strength of $t_H$ via H bonds versus $t_C$ via covalent bonds [see also (g)], as if there were breathing bonds of different lengths. (a) Reproduced from Ref. [62], (b-d) Reproduced from Ref. [70], (e-h) Reproduced from Ref. [71].

### 3.1.3. Kagome HOF monolayers on metal-vacuum interfaces

Organic molecules typically exhibit a large gap between the highest occupied molecular orbital (HOMO) and the lowest unoccupied molecular orbital (LUMO). Besides, the electronic coupling strength of hydrogen bonds is relatively weak. Therefore, HOFs are often large-band-gap insulators[72,73]. As a result, even if a kagome HOF monolayer is successfully fabricated, it is challenging to make the kagome bands cross the Fermi level. To address this issue, Perepichka, Liu and coworkers proposed a concept of donor-acceptor hydrogen-bonded organic frameworks (DA HOFs). Their first-principles calculations demonstrated that appropriately designed hydrogen-bond interactions could significantly enhance charge transfer and donor/acceptor abilities of the molecules by stabilizing/destabilizing their LUMO/HOMO levels. This designed interaction reduces the bandgaps of DA HOFs, enabling the formation of a flat band with a bandwidth of less than 0.06 eV near the Fermi level. The authors also explored the possibility that partial filling of this flat band could induce Stoner ferromagnetism[74]. Experimentally, Qi, Würthner, Haldar and coworkers, inspired by this proposal, used 3,9-diboraperylene diborinic acid derivative $B_2$-Per as building blocks (Fig. 3b inset) and deposited them on an Ag(111) surface in UHV, where they found that $B_2$-Per molecules self-assembled into a kagome lattice[70], as shown in Figure 3b. Scanning tunneling spectroscopy (STS) measurements of this system revealed an enhanced density of states at 0.33 eV above the Fermi level (Figure 3c), and the d$I$/d$V$ map at 0.36 eV (Figure 3d) indicated that this peak is, most likely, related to the kagome flat band. A tight-binding model was constructed to support that the organic monolayer corresponds to a regular kagome lattice with a hopping constant of $t$ = 0.44 eV[70].

Interestingly, researchers found that organic molecular systems can exhibit a phenomenon where the molecular structure forms a regular kagome lattice, but due to varying ease of electron transfer in different points of the lattice, the electronic band structure exhibits characteristics of a breathing kagome lattice in HOFs[71]. Pan, Liu, Li, Gao and coworkers deposited 1,3,5-tris(4-hydroxyphenyl)benzene (THPB) molecules on an Au(111) surface. These THPB molecules self-assembled into a HOF monolayer where corner benzene rings (CBRs) of THPB molecules formed a kagome lattice (Fig. 3e). Angle-resolved photoemission spectroscopy (ARPES) measurements revealed a flat band at -2.6 eV below the Fermi level over the whole Brillouin zone (Fig. 3f), which is, to the best of our knowledge, the first direct observation of a flat band in the reciprocal space within a monolayer material. Theoretical calculations identified two possible hopping pathways of electrons in this HOF monolayer, i.e. hopping within each THPB molecule through covalent bonds ($t_B$, $t_C$ in theoretical model) or among THPB molecules through hydrogen bonds ($t_H$), as shown in Fig. 3g. Largely different values for the hopping constants were revealed by fitting the results of density functional theory (DFT) calculations using a tight-binding breathing kagome lattice Hamiltonian, that are $t_C$ = 0.26 eV and $t_H$ = 0.05 eV. Thus, despite the similar lattice point distances ($d_C$ = 7.4 Å and $d_H$ = 7.2 Å), the unbalanced hopping strengths lead to the formation of a distinct electronic breathing kagome lattice (Figure 3h), which explains the observed flat band in ARPES measurements to a kagome flat band.

### 3.1.4. Kagome COF monolayers on metal-vacuum **interfaces**

In comparison to hydrogen bonding, stronger covalent bonding can enhance intermolecular electronic hybridization, delocalizing electrons across individual building-block molecules. Molecular building blocks with the $D_{3h}$[75] or $C_3$[76] symmetry were proposed for constructing kagome COF monolayers. A wide spectrum of intriguing physical properties were predicted in these monolayers by theory, including Dirac semimetals[75], high-mobility semiconductor carriers[75], and second-order topological insulators[76]. Subsequently, Contini, Perepichka, Rosei, Gallagher and coworkers successfully grew a mesoscale honeycomb-kagome lattice using tribromotrioxaazatriangulene (TBTANG) molecules (Fig. 4a inset) on Au(111) in UHV conditions (Fig. 4a). A Dirac cone (Fig. 4b) and a flat band residing at 1.8 eV below the Fermi level were observed using ARPES[77]. However, since a honeycomb lattice also contributes Dirac cones, the exact origin of the experimental observed Dirac cones remains unclear. However, due to the strong bonding in COFs, mesoscopic-scale COF kagome monolayers, as discussed above, are relatively rare. Therefore, the synthesis of large-scale COF kagome structures remains a significant challenge and requires further development.

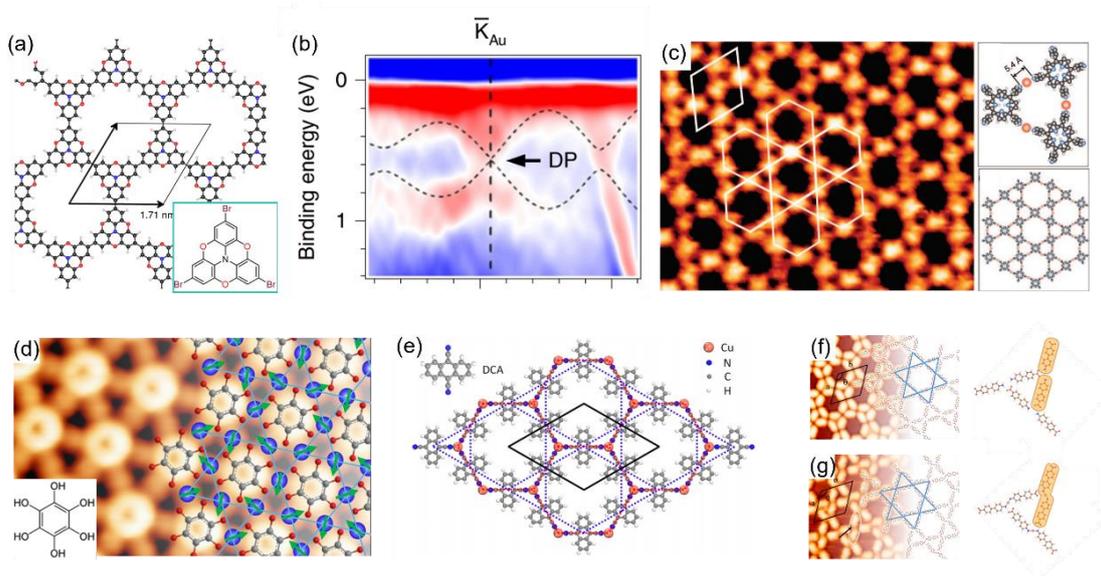

Figure 4 (a) Atomic structure of a kagome lattice constructed by TBTANG molecules (inset). (b) Second derivative plot of the spectrum for PBTANG/Au(111). (c) STM image of a kagome network formed by Au-TPyP (right). $I = 0.3$ nA, $U = -1.3$ V. The white lines indicate the unit cell and kagome symmetry. The right panel shows the structural model of the building block and the kagome network, with N, C, H, and Au atoms represented by blue, gray, white, and orange spheres, respectively. (d) STM topography image of a kagome lattice formed by benzene-1,2,3,4,5,6-hexaol (inset at bottom left) and Fe atoms, with the atomic structure overlaid on the right. Fe atoms are represented by blue spheres forming the kagome lattice, with arrows indicating the direction of the magnetic moments on the Fe atoms. $V$ = -10 mV, $I$ = 1nA. (e) Atomic structure of DCA$_3$Cu$_2$. (f-g) STM images (left) and structural models (right) of two types of kagome networks formed by ABPCA deposited on a Cu(111) surface. C, H, O, and N atoms are represented by gray, white, red, and blue spheres, respectively. $I_t$ = 100 pA, $V_b$ = -1 V. Scan area is 10 × 10 nm$^2$. (a-b) Reproduced from Ref. [77], (c) Reproduced from Ref. [78], (d) Reproduced from Ref. [79], (e) Reproduced from Ref. [80], (f-g) Reproduced from Ref. [81].

### 3.1. 5. Kagome MOF monolayers on metal-vacuum interfaces

Organic molecules aside, metal atoms were also introduced to form kagome monolayers with organic molecules through metal coordination bonds, as referred to metal-organic frameworks (MOFs). Metal coordination bonds possess relatively high cohesive energies (0.2–2.0 eV) compared with non-covalent interactions, which are reversible, and exhibit directionality and selectivity[82]. Moreover, the inclusion of metal atoms endows MOFs with a range of interesting properties, like introduction of $d$ orbitals in electron hopping[83], local magnetic moments[84,85], and so on, attracting significant attention in chemistry[82,86], physics[87,88], and other fields[89–91]. As noted in the introduction of subsection 3.1, while organic monolayers are generally easier to grow at solid-liquid interfaces, the growth of MOF monolayers or ultra-thin films remains an exception. In the liquid phase, MOFs were typically synthesized from metal salts and organic molecules, posing challenges to precisely control their thickness[82]. Layer-by-layer (LbL) assembly of MOFs[92,93], as proposed in 2007[94] and achieved to ultrathin films in 2013[93], allows to utilize nano-sized control ability over thickness in each growth cycle, but the range of applicable MOFs is limited[95]. Although this approach presents a promising direction for future exploration, the synthesis of kagome MOFs at solid-liquid interfaces has yet to be reported; this is the reason why the discussion of MOFs in 3.1.1 is missing.

Unlike at the solid-liquid interfaces, well controlled preparation of kagome MOF monolayers is much easier in UHV. In 2009, Lin and coworkers[78] synthesized the first kagome MOF monolayer by depositing 5,10,15,20-tetra(4-pyridyl)porphyrin (TPyP) molecules (Figure 4c, upper right panel) on an Au(111) surface using organic-molecular-beam-epitaxy (O-MBE). In-situ STM imaging manifests that a kagome latticed structure forms after annealing the as-deposited TPyP molecules on the Au(111) surface at 250 °C(Figure 4c, left panel). A structural model (Fig. 4c, lower right panel) was proposed for this observation, in which adjacent TPyP molecules are connected by a single Au atom (Fig. 4c upper right) forming a kagome lattice (Fig. 4c lower right) with a lattice constant of 4.1±0.1 nm. Similar experiments performed on Cu(111) and Ag(111) surfaces do not show the formation of kagome monolayer comprised of TPyP molecules, as ascribed to specific coordination properties of Au atoms by the authors.

Ni is a commonly used metal element in MOFs. The first MOF with a Ni kagome lattice was synthesized by Nishihara's group in 2013[96]. Later that year, Liu and coworkers predicted that this material could become a topological insulator under appropriate doping[97].

The metal atoms in kagome MOF monolayers can be magnetic, making the magnetism within kagome latticed MOF monolayers a topic of extensive exploration[98–101]. In particular, the magnetic atoms forming a kagome lattice could serve as an intriguing platform for studying frustrated magnetism if the nearest neighboring spin-exchange coupling of them is not ferromagnetic (FM). In line with this idea, Lin, Zhao and coworkers[79] deposited benzene-1,2,3,4,5,6-hexaol molecules (Fig. 4d inset) and Fe atoms sequentially on an Au(111) surface. A triangular lattice, formed by the benzene-1,2,3,4,5,6-hexaol molecules (bright rings), presents in a representative STM image. However, Fe atoms, appearing as rods connecting molecular rings, form a kagome lattice, as marked by blue dots in Fig. 4d right. Theoretical calculations indicated that the Fe atoms are antiferromagnetically coupled, and the system exhibits a highly degenerate ground state, providing a potential platform for studying kagome frustrated magnetism..

The strong correlation characteristics of kagome lattices can also be observed in MOFs. Schiffrin and coworkers[102] deposited dicyanoanthracene (DCA) molecules (Figure 4e inset) and Cu atoms on an Ag(111) surface, constructing a $DCA_3Cu_2$ net, where the DCA molecules formed a kagome lattice, as shown in Figure 4e. STS measurements revealed the Kondo effect within the $DCA_3Cu_2$ network. Theoretical calculations indicate that the oxidation state of Cu ions in the DCA$_3$Cu$_2$ network is +1, with 3$d$ orbitals fully occupied, so Cu itself does not exhibit a magnetic moment. The magnetic moments arise mainly from strong electron correlation effects in the kagome bands of the DCA$_3$Cu$_2$ network, where these moments are localized on the DCA molecules. The free electron gas on the Ag(111) surface interacts with the localized magnetic moments, resulting in the emergence of the Kondo effect. Besides magnetism, , kagome MOF monolayers have been predicted to host topological insulators[55], Chern insulators[103], multiple Hall effects[104], superconductivity[53,105] and among other interesting physical phenomena, which await experimental realization. Given the limited space, these are not discussed in detail here.

## 3.1.6 Multi-interaction-bonded kagome networks on metal-vacuum interfaces

Beyond a single type of bonding, combination of two or more types of bonding were employed to construct kagome molecular monolayers, providing additional thermodynamic stability[66,81,106]. For instance, Chi, Li and coworkers[81] deposited 4-aminobiphenyl-4'-carboxylic acid (ABPCA, Fig. 4f right) on a Cu(111) surface in UHV at 300K and found that ABPCA molecules formed an kagome HOFs monolayer through N-H···O hydrogen bonds (Figure 4f). However, after annealing at 450K, the amino groups of two ABPCA molecules underwent a dehydrogenation reaction, forming N-Cu-N bonds, leading to a structural phase transition and the formation of a kagome monolayer shown in Figure 4g, where both hydrogen bonds and metal−organic bonds coexist. This heating induced structural phase transition changes the bond strength and degree of hybridization, providing a simple method to tune the hopping constants between different lattice points in organic kagome monolayers.

### 3.1.7. Molecule-decoupled substrates

Over the past two decades, significant advancements have been made in surface synthesis methods and characterization techniques, leading to remarkable progress in the study of kagome organic monolayers. However, despite theoretical predictions of various intriguing physical properties in kagome organic monolayer[53,57,74], observation and validation of these properties remain challenging in experiments[70,107]. For instance, although many theoretical models predict well-defined kagome bands near the Fermi levels in kagome organic monolayers, it has always been difficult to observe these bands in monolayers supported by substrates like HOPG[108,109] or metal surfaces[66,102,110]. In rare cases, these bands were observable, but the positions of their energy levels are often far from the Fermi level[71].

The major reason for this theory-experiment discrepancy arises from unconsidered substrate-molecule interactions in most theoretical models. Theoretical models often do not consider the influence of the substrates, instead, focusing on freestanding kagome organic monolayers solely. However, extensive experimental evidence suggests that substrates significantly affect the geometric and electronic structure of these monolayers. These effects include: 1) the substrate can change the structure of the adsorbed molecular monolayer[111]; 2) electronic hybridization between the (semi)metallic substrate (e.g., noble metal (111) surfaces or HOPG) and the organic molecules can disrupt the intrinsic band structure of the freestanding kagome organic monolayer[102]; and 3) the charge doping effect of the substrate can shift band features such as flat bands away from the theoretically predicted energy positions[110].

For the reasons mentioned above, calculating the properties of freestanding organic networks to explain experimental results or predict new phenomena may be inaccurate, especially for properties sensitive to the electronic structure, such as topological properties. Even if a freestanding monolayer exhibits certain desirable properties, such as non-trivial topology, these properties can be destroyed by interaction with the substrate. This challenge has, on the theoretical side, driven further work to more thoroughly consider the impacts of substrate on the electronic structures of these monolayers. Moreover, it has, on the experimental side, motivated experimental efforts to decouple or weaken the molecule-substrate coupling. In the past few years, many attempts of the decoupling were made and significant progress has been achieved in, e.g., BN/metal surfaces, graphene/metal surfaces, and other two-dimensional materials. This subsection discusses the advancements in the growth and characterization of decoupled kagome molecular monolayers and provide an outlook on future developments.

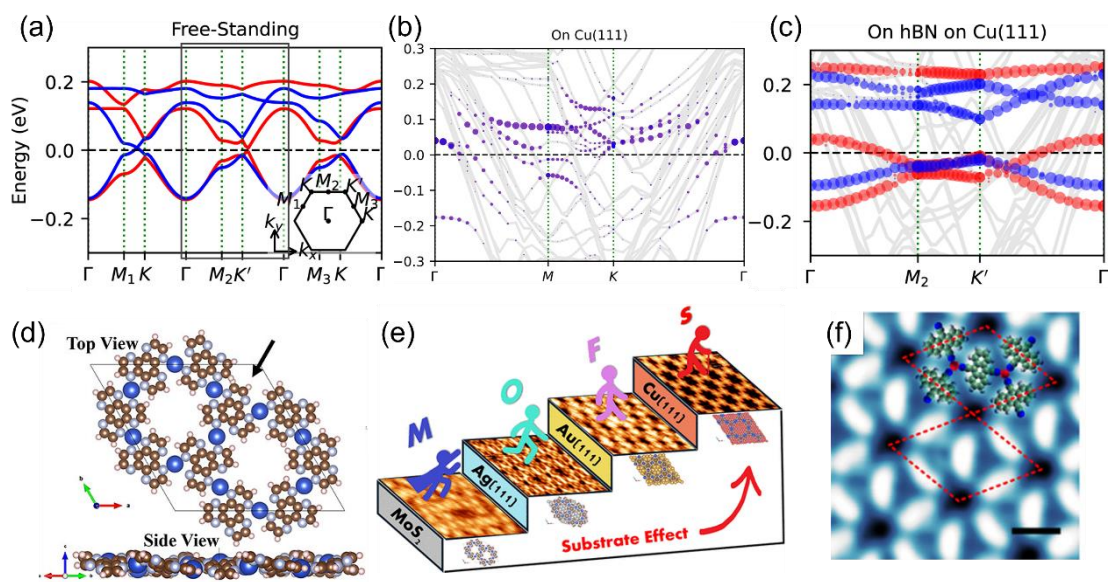

Figure 5 (a-b) Band structure of $DCA_3Cu_2$ without a substrate (a), on a Cu(111) substrate (b) and on an h-BN/Cu(111) substrate (c). In (a) and (c), the red and blue represent the spin-up and spin-down bands of $DCA_3Cu_2$. In (b), the band projection of $DCA_3Cu_2$ is shown by purple dots. (d-e) (d) Top and side views (along the black arrow) of the DFT-optimized structure of a free-standing $Cu_3HAT_2$ framework. (e) Schematic diagram of the interaction strengths of $Cu_3HAT_2$ on different substrates. (f) STM image of $DCA_3Cu_2$ on $NbSe_2$ substrate. $U = 1.0$ V, $I = 11$ Pa. (a-c) Reproduced from Ref. [112], (d-e) Reproduced from Ref. [111], (f) Reproduced from Ref.[113].

Medhekar and coworkers used DFT+$U$ calculations to study effects of different substrates on the $DCA_3Cu_2$ kagome monolayer (Fig. 4e). They found that the freestanding $DCA_3Cu_2$ monolayer is magnetic, but when placed on the Cu(111) substrate, the magnetism disappears, and the bands strongly hybridize with the substrate's bands (Fig. 5a,b). However, if a layer of h-BN is inserted between DCA3Cu2 and the Cu(111) substrate, the magnetism of $DCA_3Cu_2$ is restored. Furthermore, the band structure shows that the hybridization between $DCA_3Cu_2$ and the substrate is significantly reduced (Fig. 5c), indicating that the monolayer h-BN is effective in weakening the molecule-substrate interaction. [112]. Afterward, Schiffrin, Medhekar, Powell and coworkers successfully synthesized the DCA$_3$Cu$_2$ kagome monolayer on an h-BN/Cu(111) substrate[114]. Their STS characterization matched well with the spectral function results from dynamical mean-field theory (DMFT) calculations, proving that h-BN weakens the effect of the metal substrate. However, it is important to note that the strength of molecule-substrate interactions varies significantly depending on the specific molecules and substrates used in the process, so whether the results observed from DCA$_3$Cu$_2$ are generally applicable remains to be explored in further studies.

Another approach for decoupling lies in directly growing organic kagome monolayers on the bulk or multi-layer phase of two-dimensional vdW materials. In 2024, Lin, Huang, Shi, and coworkers[111] grew a Cu$_3$HAT$_2$ (HAT = 1,4,5,8,9,12-hexaazatriphenylene) monolayer on Au(111), Ag(111), Cu(111), and MoS$_2$ substrates. By comparing the lattice constant of the freestanding monolayer and STM simulation images with the experimental results, they obtained structures of Cu$_3$HAT$_2$ monolayers on different substrates and summarized a hierarchical diagram (Fig. 5e) showing the varying magnitudes of the effects of different substrates on the monolayer structure. They found that the framework grown on MoS$_2$ is nearly identical to its freestanding counterpart in terms of structure, suggesting that wo-dimensional vdW can serve as weakly coupled platforms for growing organic monolayers. There has also been progress in growing kagome monolayers on two-dimensional materials. In 2021, Liljeroth, Yan, and coworkers[113] grew the DCA$_3$Cu$_2$ kagome MOF monolayer, mentioned earlier, on a two-dimensional vdW superconducting material, NbSe$_2$, STM image shown in Fig. 5f. The growth of kagome monolayers on two-dimensional materials is still in its early stages, and warrants further research.

The synthesis of organic kagome networks, involving reactions that require bond cleavage and reformation, is less efficient on two-dimensional material substrates due to their low catalytic efficiency[115,116]. For systems with stronger intermolecular interactions and hybridization, such as COFs or MOFs formed through strong bonds, further exploration is needed to achieve synthesis on vdW material surfaces. Alternatively, easily transferable COFs were recently synthesized at the gas-liquid interface[117,118], suggesting a potential solution: first grow strongly bonded kagome organic monolayers at a gas-liquid interface, then transfer them onto an insulating substrate for characterization.

## 3.2. Kagome-structured surface states

Shortly after the advent of STM in the last century, scientists began using it to manipulate atoms on metal surfaces, thereby indirectly controlling the surface electronic states of the metal[119]. For instance, quantum confinement effect of the Shockley surface state of a Cu(111) surface was observed in 1993 by M. F. Crommie, C. P. Lutz and D. M. Eigler, in which a quantum corral was artificially built by manipulating Fe atoms[120]. Beyond individual atoms, organic molecules were recently used as repulsive potential barriers to confine surface electronic states on metal surfaces, leading to the formation of various electronic kagome lattices[121]. Here, this subsection introduces several examples of electronic kagome lattices constructed using organic molecules.

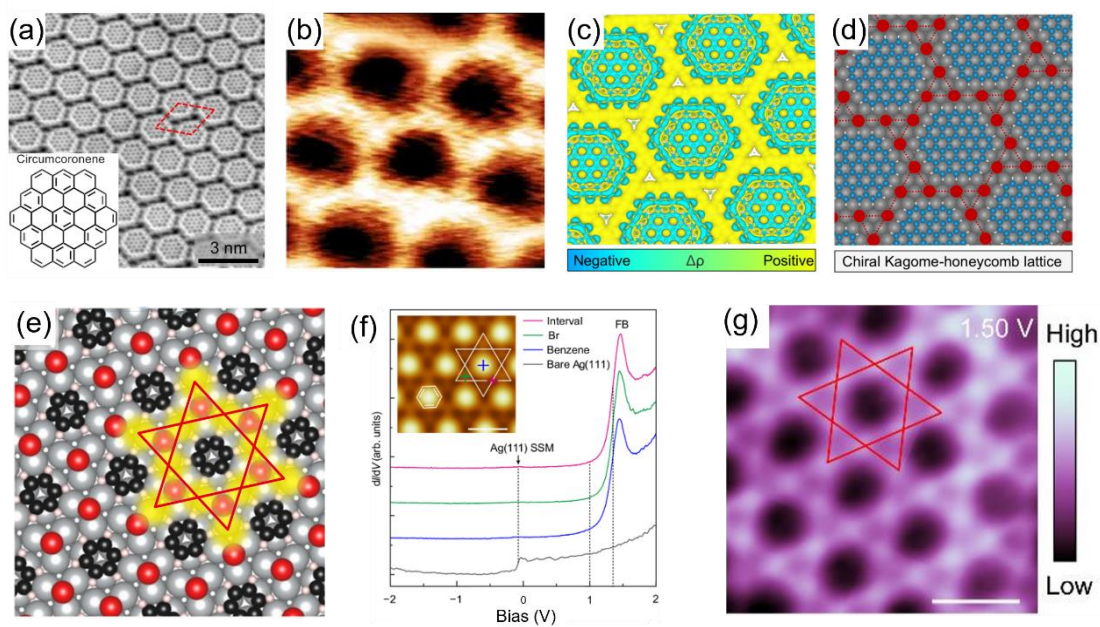

Figure 6 (a-d) Non-contact atomic force microscopy image of a superlattice formed by circumcoronene (inset) (a), d$I$/d$V$ map at 1.15 V (b), electrostatic potential (c), and schematic of the chiral diatomic kagome lattice (d). (e-g) Structural model of the structure of benzene/Br/Ag(111) (e), d$I$/d$V$ spectra (f), and d$I$/d$V$ map at 1.50 V (g). The inset in (f) shows the STM topography image, with the colored-cross indicating the STS measurement position.(a-d) Reproduced from Ref. [40], (e-g) Reproduced from Ref. [38].

In 2021, Lu, Wu, Jelínek and coworkers[40] devise a strategy for the ultrahigh-yield synthesis of circumcoronene molecules on Cu(111) via surface-assisted intramolecular dehydrogenation of the rationally designed precursor, followed by methyl radical-radical coupling and aromatization. They observed that these molecules self-assembled into a triangular lattice on the Cu(111) surface (Figure 6a). Two peaks residing at 0.36 V and 1.15 V were identified in a d$I$/d$V$ spectrum acquired at the interstitial regions among the molecules, which cannot be explicitly ascribed to tunneling into any frontier molecular orbitals. Figure 6b shows a d$I$/d$V$ map acquired at 1.15 V, revealing that this state is primarily distributed in the interstitial regions, while the map acquired at 0.36 V suggests an identical feature. This feature suggests that the two peaks originate from the surface electronic states of Cu(111). Electrostatic potential calculations (Figure 6c) revealed that the circumcoronene molecules exhibit negative potentials to electrons, confining the surface electrons within the interstitial regions. This confinement of surface electronic states leads to the formation of an electronic chiral diatomic kagome lattice (Figure 6d) and the emergence of two kagome flat bands.

Non-covalent interactions were recently introduced to construct electron repulsive barriers on metal surface. On-surface synthesis of halogen hydrogen-bonded organic frameworks (XHOFs) was recently proposed by Wang, Ma and coworkers [38] as a general strategy to construct electron repulsive barriers for realizing electronic kagome lattices. They successfully realized regular, breathing, and chiral diatomic electronic kagome lattices on brominated Ag(111) and Au(111) surfaces. For instance, Fig. 6e shows the atomic structure model of a benzene/Br/Ag(111) superlattice. The d$I$/d$V$ spectra acquired on the modified surface identify a peak residing at 1.5 V. Its

d*I*/d*V* mapping image (Figure 6g) clearly displays a distinct regular kagome lattice pattern in the spatial distribution of this state.

Electronic kagome lattices, constructed using surface electronic states confined by molecular and/or atomic adsorbents, broadens the spectrum of molecules that can be employed, without necessitating the molecular monolayers to have a kagome topology. However, this approach also introduces new challenges. Due to the complex interactions between the molecules and the substrate, even with knowledge of the properties of the molecules, it is not easy to determine, before an experiment, whether placing the molecules on the substrate will confine the metal surface electrons to form an electronic kagome lattice. Additionally, theoretical calculations must thoroughly consider molecule-substrate interactions, such as molecule-substrate alignments, interfacial electronic and mechanic couplings, all of which add to the complexity and computational burden. As a result, this method is still in its exploratory stages and requires further in-depth research.

# 4. Inorganic kagome lattices on surfaces

In addition to organic molecules, atoms or inorganic molecules can also be employed to construct kagome monolayers on surfaces. This section introduces the advances in this area, including the construction of atomic kagome monolayers through templating, and the formation of electronic kagome lattices via atomic manipulation, deposition, and orbital hybridization on surfaces.

## 4.1. Atomic kagome lattices constructed via surface templates

A kagome-shaped array of potential wells on a surface can be utilized through a specific surface structure. These wells further capture deposited atoms and arrange them into a kagome lattice, allowing for the controlled formation of a single-atom-thick kagome monolayer. At this stage, the specific surface structure acts as a template for constructing a kagome monolayer on the surface. This concept has already been realized by depositing K atoms on a blue phosphorus-gold alloy[29,122].

Figure 7a shows an atomic structure of a blue phosphorus-gold alloy, $P_2Au$, where two adjacent hexagons (shadowed hexagons) provide two potential wells capable of accommodating atoms. As shown in Fig. 7b, these potential well pairs position on centers of connecting lines of the nearest-neighbor points in triangular lattices, so form kagome lattices[123,124]. If one of the two adjacent potential wells is occupied, it is possible to construct a regular or twisted kagome lattice[29]. Du, Zhou, Hao, Zhuang and coworkers synthesized the blue phosphorus-gold alloy $P_2Au$ by depositing P atoms on an Au(111) surface, then depositing K atoms to occupy the potential wells (red dots in Figure 7c), thereby constructing a twisted kagome lattice shown in Figures 7c-d. STS measurements revealed a peak in the d$I$/d$V$ spectra, possibly originating from the flat bands of the twisted kagome lattice.

When the two adjacent potential wells merge into one (Figure 7e), they form a template for constructing a kagome lattice (indicated by the solid red lines in Figure 7e). Chen, Li, Zhang and coworkers achieved the growth of this single-well blue phosphorus-gold alloy $P_4Au_3$ by first depositing K atoms on an Au(111) surface under UHV conditions, resulting in the functionalization of the Au(111) surface with K atoms, followed by the deposition of P atoms[122]. STM results (Figure 7f) showed that the K atoms were located within these wells, forming the expected kagome structure.

This template-assisted construction of kagome monolayer on surfaces offers a wide range of flexibility, allowing various parameters within the kagome monolayer to be tuned by selecting different template shapes, sizes, and materials. However, at the current stage of research, the available templates for the construction remain relatively scarce, limiting a broader application of this approach. Therefore, further development is needed to explore and identify new templates capable of forming kagome potential energy surfaces.

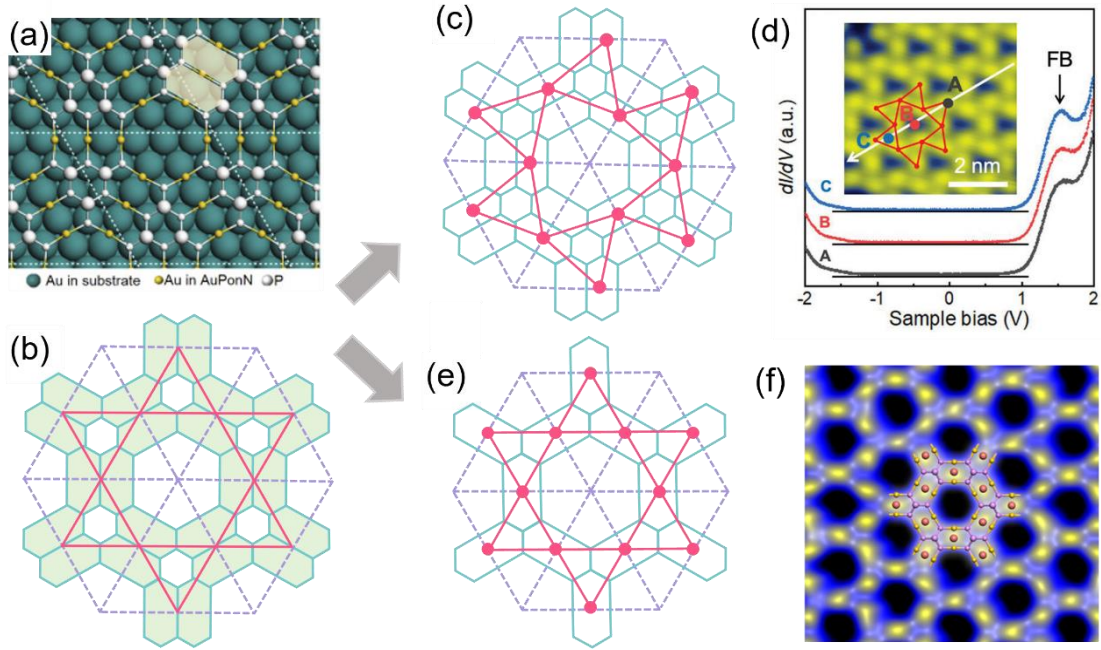

Figure 7 (a) Atomic structure of $P_2Au$ on Au(111). (b) Schematic of the arrangement of the potential wells in $P_2Au$. The shaded hexagons represent potential wells. The purple dashed lines indicate connecting lines of the nearest neighbor points in triangular lattices. The red solid lines highlight the kagome shape. (c) Schematic of a twisted kagome lattice constructed from the double-potential well blue phosphorus-gold alloy $P_2Au$. (d) Differential conductance spectra of the twisted kagome lattice shown in (c), with the spectra vertically offset for clarity. The inset shows an STM image with A, B, and C marking the STS measurement positions. $V_s$ = -0.1 V, $I$ = 200 pA. The underline beneath each spectrum indicates the zero density of states level. (e) Schematic of the single-potential well blue phosphorus-gold alloy $P_4Au_3$. (f) STM images of potassium-atom-adsorbed $P_4Au_3$, with the atomic structure overlaid. $V_s$ = -0.02 V. (a) Reproduced from Ref. [124], (d) Reproduced from Ref. [29], (f) Reproduced from Ref. [122].

## 4.2. Atom-scale electronic kagome lattices

Atomic manipulation is a powerful tool for assembling atoms in finite sized lattices. By using this tool, researchers have successfully constructed a series of artificial electronic lattices on Cu(111) surfaces, including honeycomb lattices[125,126], Lieb lattices[127], and kagome lattices[128]. In 2019, Morais Smith, Swart and coworkers[128] used a STM tip to manipulate CO molecules on a Cu(111) surface, arranging them into the structure indicated by the black dots shown in Figure 8a inset. The polar CO molecules induce localized surface potential wells, thereby repulsing the Cu surface states from the adsorption sites. These repulsive sites effectively confine the surface electronic states into a finite-sized breathing kagome lattice, as shown in Figure 8a. They further demonstrated that this finite sized lattice is a second-order topological insulator, exhibiting stable zero-energy corner modes (Figure 8b).

Flexibility of the most pronounced advantage for constructing lattice using tip-based atomic or molecular manipulation. Customizable type of lattice and strength of inter-site hopping are crucial for exploring novel quantum effects. However, this method is typically inefficient and lacks scalability. Otte and coworkers[129] provide an effective solution to the scalability challenges of atomic manipulation. They developed a kilobyte rewritable atomic-scale memory consisting of $10^4$ atoms, based on the self-assembly of Cl atoms on a Cu(100) surface and STM manipulation. However, a subsequent challenge arises: the remaining metal surface states may hybridize with the kagome bands or interfere with their observation. Atomic manipulation on surfaces of semiconductors or insulators could be a likely solution for constructing neat surface electronic kagome lattices[130,131]. In 2024, Fölsch and coworkers[132] built In quantum dots using atomic manipulation on the (111) surface of a InAs semiconductor surface, on which a peak, arising from individual quantum dots consisting of six In atoms, was observed at -75 mV in the d$I$/d$V$ spectra shown in Figure 8c . Therefore, atomic manipulation on surface of semiconductor and insulator could offer a route toward construction of neat kagome bands near the Fermi level. Additionally, manipulating magnetic atoms on metal[133] or insulator[134] surfaces with an STM tip could also be a promising approach for constructing kagome magnets on surface. Nevertheless, all these approaches share the same challenges of efficiency and scalability.

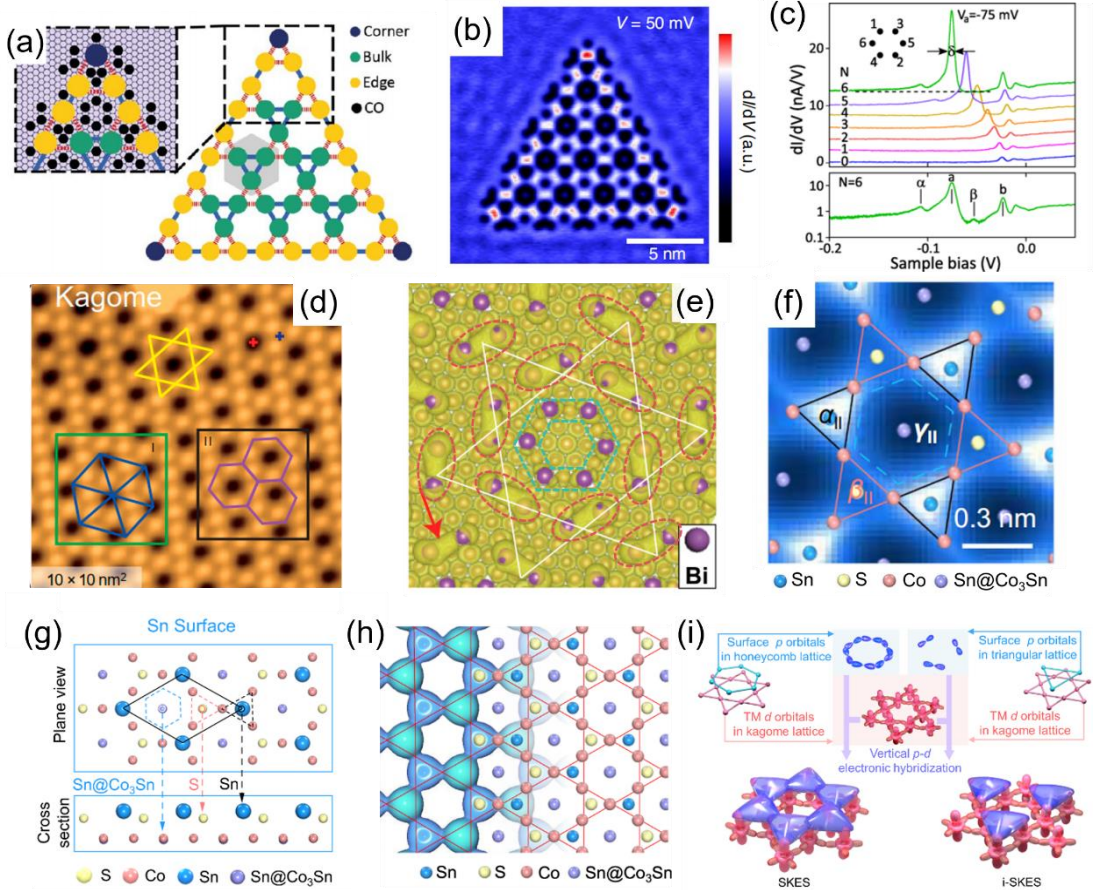

Figure 8 (a-b) (a) Schematic of the kagome lattice formed by CO on a Cu(111) surface (inset) and (b) d$I$/d$V$ map at 50 mV bias. (c) Upper panel: size-dependent d$I$/d$V$ spectra (set point: 1 nA, 0.3 V; $V_{mod}$ = 1.8 mV) with $N$ representing the number of adatoms taken at the center of the In hexagon (assembly sequence highlighted by the inset): a strong peak evolves at $V_a = -75$ mV for the complete hexagon ($N= 6$) together with smaller peaks at higher and lower energies; the width $\delta$ of the main peak (full width at half maximum) is measured relative to the dashed horizontal baseline. Lower panel: logarithmic plot of the $N = 6$ spectrum; peaks denoted α and β are replicas of peaks a and b induced by inelastic electron tunneling. (d) STM image of the kagome lattice formed by Bi on the Au(111) surface. $V_s$ = 1.5 V, $I_t$ = 1000 pA. (e) Atomic structure model of Bi/Au(111) surface and the electronic density distribution. The red and cyan dashed lines outline the Type I and II Bi atoms. (f-i) (f) Chemical-bond-resolved nc-AFM image of the Type-II surface in $Co_3Sn_2S_2$, the Sn-terminated surface. Three distinct regions within a unit cell with bright, blurry, and dark contrast, which are marked by black solid line triangles, red solid line triangles, and a blue dashed line hexagon, are labeled as $α_{II}$, $β_{II}$, and $γ_{II}$ regions. The atomic structure superimposed is the Sn surface with the underlying S and $Co_3Sn$ plane. (g) DFT optimized surface structure on the Sn surface. (h) Isosurface contour of $|\psi|^2$, integrated from −0.38 eV to the $E_F$, on the Sn surface superimposed with the atomic structure of the three topmost atomic layers of the Sn surface in the right part. Red solid lines highlight the kagome pattern. (i) Schematic of SKES (the left part) and i-SKES (the right part) formation through vertical $p$–$d$ hybridization. (a-b) Reproduced from Ref. [128], (c) Reproduced from Ref. [132], (d-e) Reproduced from Ref. [135], (f-i) Reproduced from Ref. [136].

When atoms are directly deposited onto a surface, their interactions with each other and with the substrate can also lead to the formation of electronic kagome lattices, which include states associated with adatoms and potentially surface states. In 2019, Han and coworkers[137] studied the coverage dependent evolution of atomic and electronic structures of self-assembled Bi adatoms on an Au(111) surface. As the Bi coverage reaches 0.63 ML, a typical STM image of the Bi adatoms on the surface exhibited a kagome latticed pattern[137], which was also observed in another STM study[135] (performed by Qin, Zhang and coworkers) on the same surface, as shown in Figure 8d. Density functional theory calculations by Qin et al. reveal a corresponding atomic structure for the surface (Figure 8e), in which the Bi atoms are categorized into two types. Type-I Bi atoms (outlined by red dashed lines) aggregate into dimers with a higher density of electronic states, which appear as bright spots located at the midpoint of the dimers in STM images acquired at certain bias voltages, forming a kagome lattice. Type-II Bi atoms (outlined by cyan dashed lines) resides inside the hexagons of the kagome lattice and appears as dark voids in STM images. The Bi adatoms do not form a kagome lattice in atomic arrangement; instead, the kagome pattern observed in STM images reflects an electronic kagome lattice arising from the electronic states on the Bi dimers (Figure 8e). This approach to constructing electronic kagome lattices by depositing atoms onto a surface is straightforward and require no further manipulation. However, it depends strongly on the choice of atoms and substrates, and currently neither experiments nor theory can predict which systems will yield electronic kagome lattices. Thus, establishing a general construction strategy, possibly empowered by machine learning models, remains an open question for future exploration.

Kagome latticed electronic states can also be constructed through interlayer orbital hybridization between a honeycomb-latticed surface layer and a kagome layer underneath, as referred to surface kagome electronic states (SKES). This strategy was recently proposed by Gao, Ji, Yang and coworkers[136] and developed through their studies on $Co_3Sn_2S_2$, a kagome magnet, surfaces. Two types of $Co_3Sn_2S_2$ surfaces were observed in STM experiments, which are terminated by a S (the S-terminated surface) and a Sn (the Sn-terminated surface) atomic layer, respectively. The triangular Sn surface, however, exhibits properties related to the kagome lattice, such as Weyl fermion arcs, which are not present on the S-terminated surface. The qPlus non-contact atomic force microscopy (nc-AFM) also images a kagome latticed pattern on the Sn-terminated surface, as shown in Figure 8f. The DFT calculations reveal that the height difference between the Sn atomic layer and the underlying S atomic layer is only 0.56 Å in the Sn-terminated surface (Figure 8g, side view). These two atomic layers together form a honeycomb lattice (Figure 8g, top view), whose lattice points are directly positioned above the centers of the $Co_3$ triangles in the $Co_3Sn$ kagome layer underneath. Strong electronic hybridization occurs between the *p* orbitals of the surface Sn and S atoms in the honeycomb lattice and the *d* orbitals of Co atoms in the $Co_3Sn$ kagome layer, effectively imprinting the triangular-shaped electronic density of the $Co_3$ trimers onto those of the surface Sn and S atoms. Subsequently, the *p* orbitals of Sn and S undergo in-plane hybridization, resulting in the formation of a surface electronic kagome lattice on the Sn-terminated surface, as shown in Figure 8h. On the S-terminated surface, only an incomplete SKES (i-SKES) can form because the surface consists solely of a triangular lattice of S atoms, as shown in the right panel of Figure 8i.

Inspired by these findings, thy proposed a universal strategy for constructing SKES, as illustrated in the left panel of Figure 8i: (i) the surface and subsurface (if any) atoms fit in a honeycomb lattice, (ii) their in-plane states vertically hybridize with that of the kagome sublattice underneath, and (iii) vertically hybridized states then subsequently hybridize laterally to form an SKES, as illustrated in Fig. 8i. The universality of this strategy was further verified through DFT calculations, in which SKESs are identified on modified Sn-terminated $Co_3Sn_2S_2$ surfaces by substituting Sn with other group-13 to group-15 elements or replacing S with Se or Te atoms. In addition to the honeycomb-kagome stacking on the surface giving rise to SKES, Zhang and coworkers[138] demonstrated through a tight-binding model that the stacked honeycomb and twisted kagome lattices also exhibit kagome bands. They swept through the Inorganic Crystal Structure Database (ICSD) and discovered 298 experimentally synthesized new ideal topological materials with this stacking arrangement.

# 5. Moiré kagome few-layers

When two or more two-dimensional layers are vertically stacked with a lattice mismatch[139,140] or rotational misalignment (twist angle)[141,142], they form a larger-scale periodic structure due to the interference of their atomic arrangements, known as a moiré superlattice[143,144]. Moiré superlattices generally feature larger lattice constants, typically from few to tens of nanometers, and thus reduced inter-site hopping, thereby amplifying the effects of electron correlations. The enhanced correlation effects lead to the emergence of strong correlations or/and topological properties that are absent in the parent two-dimensional materials. These effects or properties emerge through mechanisms such as Brillouin zone folding[145,146], structural relaxation[147–149], and strong interlayer coupling[150,151]. Various intriguing physical phenomena have already been reported in moiré bilayers, such as superconductivity and correlated insulators[141,152,153], moiré excitons[154–156], ferromagnetism[157,158], and the quantum anomalous Hall effect[159,160]. Moreover, the significantly increased real-space periodicity effectively reduces the electron density required for tuning band occupancy using electrical gating, thereby expanding the range of tunability. Consequently, realizing a two-dimensional kagome lattice within a moiré bi- or multi-layer could enable the construction of a variety of novel quantum states with enhanced tunability, potentially uncovering new and exotic physical phenomena.

Twisted multilayer silicene is one of the earliest examples of building electronic kagome lattices in moiré superlattices[161]. As shown in Figure 9a, Du, Hu, Chen and coworkers discovered that STM images of a twisted multilayer silicene with a 21.8° twisting angle exhibit a kagome lattice pattern with a period of 1.7 nm. In the associated

d*I*/d*V* spectra, two narrow peaks were observed at 1.32 eV and 1.7 eV above the Fermi level (Figure 9b). The d*I*/d*V* mapping image at 1.32V clearly shows a kagome latticed pattern (Figure 9c), indicating the formation of an electronic kagome lattice and a kagome flat band at 1.32V, characterized by the localized but extended electronic states due to destructive quantum interference.

In addition to the twisted silicene, electronic kagome lattices have also been observed in other twisted moiré superlattices[162,163]. For instance, similar electronic kagome latticed patterns were observed with STM in twisted bilayer WSe$_2$ with a 5.1° twisting angle[163] (Figure 9d) and twisted bilayer graphene (BG) [162] with twisting angles of 1.07°, 0.98°, and 0.88°.

Atomic kagome lattices have also been observed in twisted moiré systems. In a twisted graphene trilayer with twisting angles $\theta_{12} = 0°$ and $\theta_{23} \cong 0.06°$, Yoo, Son and coworkers[164] observed kagome-shaped domains using dark field transmission electron microscopy (DF TEM), as shown in Figure 9e[164]. In these domains, the ABC and ACB stacking orders (atomic structures shown on the right side of Figure 9e) form the triangles of the kagome lattice, with alternating bright and dark contrasts in the DF TEM images. However, the lattice constant of this kagome lattice reaches several hundred nanometers. As a result, the hopping between kagome lattice points is negligible, and it is unlikely that this structure will exhibit kagome properties associated.

The relatively larger lattice constants of moiré superlattices make it feasible to use microscopy techniques with spatial resolutions lower than the atomic scale. One promising technique lies in scanning microwave impedance microscopy (MIM) [165–167]. In MIM, microwave signals (1–10 GHz) are directed through a sharp metal tip onto the sample, and the reflected signals are analyzed to measure the admittance between the tip and sample. The imaginary component, MIM-Im, correlates with sample conductivity, enabling local conductivity mapping. Zettl, Wang and coworkers [168] used an ultrahigh-resolution implementation of scanning microwave impedance microscopy (uMIM) to image a twisted BG/BG/hBN multilayer (inset of Figure 9f). They found that, at a twisting angle of 0.6°, the low-pass filtered MIM-Im image revealed bright spots, corresponding to high-conductivity domains, arranged in a kagome lattice pattern, as shown in Figure 9f. However, since MIM is sensitive to changes in both atomic structure[168] and electronic structure[166], it remains unclear whether the observed kagome lattice corresponds to an atomic or electronic kagome lattice.

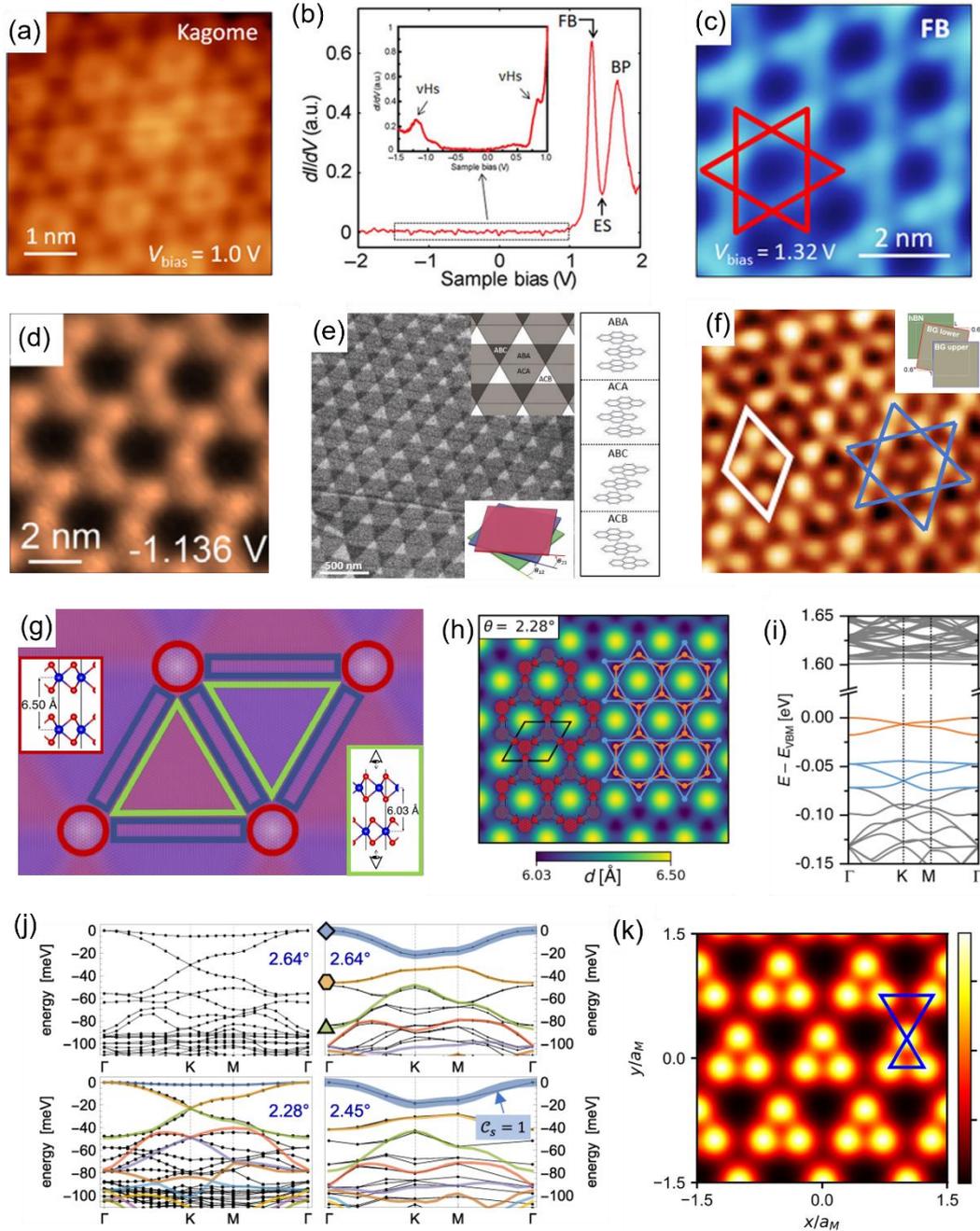

Figure 9 (a-c) STM image (a), d$I$/d$V$ spectra (b), and d$I$/d$V$ map (c) at a bias of 1.32 V of multilayer twisted silicene. In (a), $V_s$ = 1 V, $I_t$ = 100 pA, scan area 5×5 nm$^2$. (d) d$I$/d$V$ map of 5.1° twisted bilayer WSe$_2$ at a bias of -1.136 V. (e) DF TEM image of triple-layer twisted graphene with twist angles $\theta_{12}$ = 0° and $\theta_{23}$ ≅ 0.06° (inset at bottom right). The inset at top right shows the stacking configurations corresponding to the triangular and hexagonal domains of the kagome lattice, with the atomic structure diagrams of different stacking configurations on the right. (f) Low-pass filtered image of the uMIM scan of BG/BG/hBN (inset). (g-h) (g) Fully relaxed atomic structure of twisted bilayer MoS$_2$ with a twist angle $\theta$ = 1.05°. The red circles highlight the $R_h^h$ stacking regions (inset on the left), the green triangles highlight the $R_h^M$ stacking regions viewed from above or below (inset on the right), and the blue rectangles highlight the solitons. (h) Interlayer separation landscape with a schematic visualization of (left overlay) the structural elements and (right overlay) the superlattices formed by the structural elements in twisted bilayer MoS$_2$ with a twist angle $\theta$ = 2.28°. The $R_h^M$ stacking domains form a honeycomb lattice (orange) while the midpoints of solitons form a kagome lattice (blue). (h) Electronic band structure of twisted bilayer MoS$_2$ (twist angle $\theta$ = 2.28°), calculated at the DFTB level. The bands corresponding to the honeycomb and kagome lattice are highlighted. (j) Ab initio band structure of twisted bilayer ZrS$_2$, with (right column) and without (left column) spin-orbit coupling, with the top-most valence bands arising from the emergent Kagome lattice. The solid colored lines represent the band structure fitted to the topmost valence band. The thick blue line indicates a topological band with a Chern number of 1. (k) Real-space electron density of a Wigner molecular crystal in a breathing kagome lattice. (a-c) Reproduced from Ref. [161], (d) Reproduced from Ref. [163], (e) Reproduced from Ref. [164], (f) Reproduced from Ref. [168], (g-i) Reproduced from Ref. [169], (j) Reproduced from Ref. [170], (k) Reproduced from Ref. [171].

Although experimental results for kagome lattices in twisted moiré systems are still relatively limited, theoretical predictions suggest that kagome lattices can be realized across a range of moiré systems.

Heine, Kuc and coworkers[169] conducted a comprehensive theoretical study on the structural and electronic evolutions in twisted bilayer MoS$_2$ with the twisting angle varying from 0.2° to 59.6°. The fully relaxed structure of twisted bilayer MoS$_2$ is shown in Figure 9g and includes three distinct regions: stacking $R_h^h$ (red circles), stacking $R_h^M$ stacking (green triangles), and solitons (blue rectangles). In 2.28° twisted bilayer MoS$_2$, fully relaxed using a reactive force field, the $R_h^h$ stacking region (yellow bright spots in Figure 9h) forms a triangular lattice, while the centers of the solitons (blue spots in Figure 9h) constitute a kagome lattice. The lowest-order density-functional based tight-binding (DFTB) indicates that a clean kagome band structure can be found just 0.05 eV below the Fermi level, as shown by the blue bands in Figure 10i[169]. Similarly kagome bands below the Fermi level were predicted in twisted bilayer 1T-ZrS$_2$ at twist angles of 2.45°, 2.28°, 2.64°, and 3.15° (Fig. 9j) by Rubio, Kennes, Claassen and coworkers[172], using a semiconductor moiré continuum model[172] and ab initio calculations. These bands are contributed by the degenerate $p_x$ and $p_y$ orbitals of sulfur[170]. Charge neutral bilayers asides, Reddy, Devakul, and Fu[171] used a semiconductor moiré continuum model to predicted that at filling factor $n = 3$, the Coulomb interactions within each three-electron moiré site lead to a three-lobed "Wigner molecule". When these molecules are comparable in size to the moiré period, they arrange into an emergent distorted electronic kagome lattice due to the balance between Coulomb interactions and the moiré potential, as shown in Figure 9k [171]. These calculations offer general predictions for semiconductor moiré systems. Wang, Crommie, Fu and coworkers[173] observed Wigner molecule crystals using STM in a twisted bilayer $WS_2$, but the molecule size was smaller than the moiré period, preventing formation of a kagome lattice[174]. Similarly, a Wigner crystal forms in twisted bilayer graphene with a twisting angle of approximately 1°[173], where the lattice type varying by electron filling. At a filling of 3/4, a kagome latticed Wigner crystal may emerge. Beyond from twisted moiré systems, Kong, Ji and coworkers predicted that applying different in-plane strains to an untwisted graphene homo-bilayer could also yield a twisted kagome structure and kagome bands[175].

Despite encouraging progress in both experimental and theoretical research on kagome moiré lattices, several key challenges remain. Theoretically, for moiré systems with small twist angles, first-principles calculations often involve handling more than $10^3$ or even $10^4$ atoms, which pose substantial challenges for structural relaxations. This complexity limits the accuracy of structural and electronic property predictions in kagome moiré lattices, thereby limiting the ability to effectively simulate, predict, and interpret kagome lattices in these contexts. Experimentally, twisted moiré superlattices often encounter stability issues, as their twisting angle may shift under thermal, mechanical, or other forms of perturbations[176–179]. Moreover, although electronic kagome lattices are realized and kagome flat bands are observed in moiré superlattices, research into the properties of these kagome bands is still insufficient, and the ease of tuning band occupancy through electrostatic gating remains underutilized. Nevertheless, these challenges underscore valuable directions for future research. Exploring how to fully exploit the advantages of moiré superlattices and the novel physical phenomena arising from the interactions between moiré and kagome lattices represents a promising and exciting frontier for further investigation.

# 6. Mirror twin boundaries as superatoms for constructing kagome monolayers

In monolayer transition metal dichalcogenides (TMDs), grain boundaries, as a type of structural defect, significant influence the thermal, mechanical, electrical, and other properties of TMDs[180–184]. A notable example lies in the atomically sharped mirror twin boundaries (MTBs), which form between regions with a 60° relative orientation. To date, three major MTB structures, i.e. 4|4P[180] (Figure 10a), 4|4E[180], and 55|8[185], have been reported in various TMD materials. Each structures exhibits distinct electronic properties, but they all introduce relatively isolated in-gap states within the TMD bandgap.

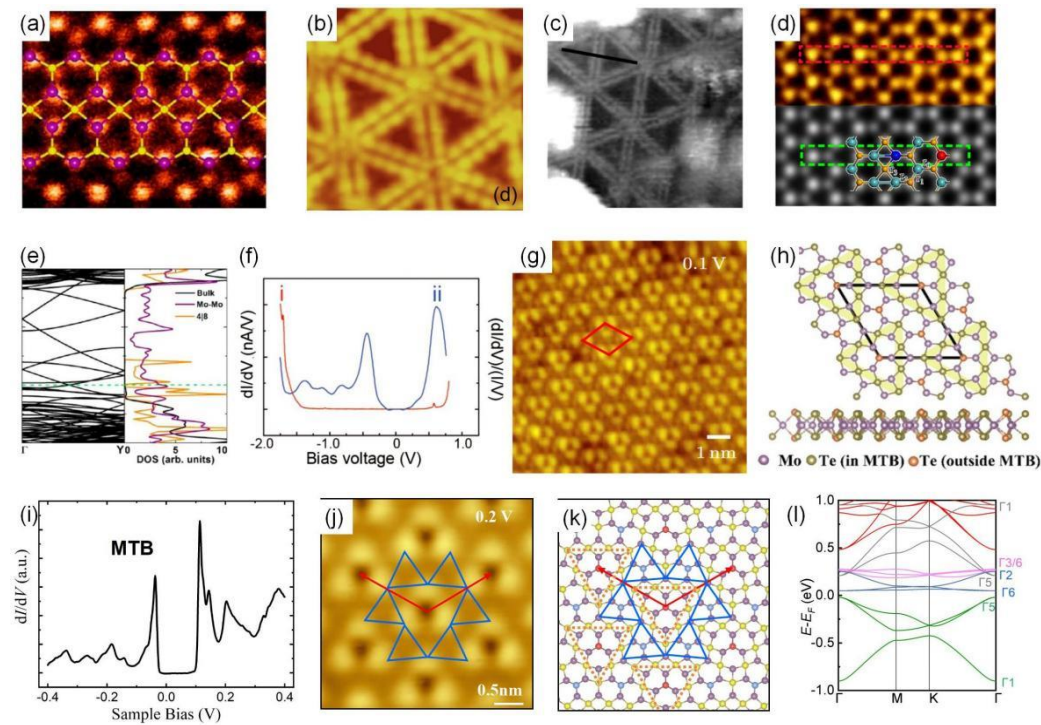

Figure 10. (a) Atomic-resolution annular dark-field STEM (ADF-STEM) images of 4|4P grain boundaries in MoS$_2$, overlaid with atomic structure diagrams. (b) STM image of MoSe$_2$ on HOPG. Size: 13 × 13 nm$^2$, $V_{simple}$ = 1.46 V. (c) STM image of MoSe$_2$ on MoS$_2$. Size: 23 × 23 nm2, $I_t$ = 0.20 nA, $V_{simple}$ = -0.78 V. (d) Experimental and simulated ADF images. Scale bar: 0.5 nm. Partial atomic structure diagrams are overlaid.(e) Band structure and local density of states (LDOS) for the 4|8 grain boundary. (f) $dI/dV$ spectra for monolayer (ML) MoSe$_2$ at the interior of the grain boundary (i) and on the MTB triangular loop (ii). (g) 2√3×2√3 superlattice in MoTe$_2$, $I$ = 100 pA. (h) Atomic structure model of Mo$_5$Te$_8$, with the side view shown below. The MTB is marked by yellow triangles. Mo atoms are represented by purple spheres, Te atoms shared by the MTB are shown in green, and Te atoms outside the MTB are shown in orange. (i) The $dI/dV$ spectra on the MTB. (j) STM image of Mo$_5$Te$_8$ at a bias of 0.2 V. The coloring-triangular (CT) lattice is highlighted by solid blue lines, and the lattice vectors are marked by red arrows. (k) Atomic structure of Mo$_5$Te$_8$, with the MTB highlighted by orange dashed lines. (l) Band structure of Mo$_5$Te$_8$. The irreducible representations of the four CT bands at the Γ point are listed on the right, marked in green, blue, purple, and gray. (a) and (e) Reproduced from Ref[180], (b) and (f) Reproduced from Ref[186], (c) Reproduced from Ref[187], (d) Reproduced from Ref[188] (g) Reproduced from Ref[189], (h) Reproduced from Ref[190], and (i-l) Reproduced from Ref[24].

In 2013, Zhou and coworkers identified two MTBs in 1H-MoS$_2$, composed of four-fold rings that share sulfur atoms, known as 4|4P and 4|4E[180]. The lattices on the both sides of the 4|4P boundary are mirror-symmetric, whereas that of the 4|4E boundary requires an additional translation operation. In 2014, Xie and coworkers grew MoSe$_2$ samples on a HOPG substrate using MBE[186]. They reported that the MTBs form a characteristic triangular inversion region, which combines into a wheel-like pattern[186] (Figure 10b). This structure was proposed to be a result of MTBs sharing, rather than the previously assumed moiré interference effect[187,191–194] (Figure 10c).. Since then, this MTB chains [191,194,195] and/or triangular loops [196–198] have been discovered in various MoX$_2$ (X = S, Se, Te) materials. In 2017, Ji, Jin, Xie and coworkers, using high-resolution ADF-STEM imaging combined with DFT calculations, confirmed that the MTB triangular rings in monolayer MoSe$_2$ correspond to 4|4P MTBs sharing Se atoms[188] (Figure 10d). As shown in Figure 10e, theoretical calculations predict that these MTBs form one-dimensional metallic quantum wires, with electronic states crossing the Fermi level. This metallic feature was experimentally verified the metallic feature at the MTBs within the wheel-like pattern. These in-gap states are typically attributed to the formation of charge ordering, such as Peierls-type charge density waves (CDW)[195,199,200] or Tomonaga-Luttinger liquids[201,202].

The smallest 4|4P MTB loops require the addition of three extra Mo atoms[203] which was experimentally realized in H-MoTe$_2$ as a prototype a $2\sqrt{3} \times 2\sqrt{3}$ superlattice on a 2H-MoTe$_2$ sample was reported by Zhang, Wang and coworkers in 2017[200] (Figure 10j), although its exact periodicity was in debate that some others initially suggesting a 2×2 super-periodicity[189] (Figure 10k). Later in 2020, Xie, Jin and coworkers[190], using scanning transmission electron microscopy, definitively identified this structure as a new layered transition metal chalcogenide, Mo$_5$Te$_8$, featuring a $2\sqrt{3} \times 2\sqrt{3}$ superlattice and consisting of the smallest MTB loops, as illustrated in Figure 10l. Three years later, Ji, Wang, Cheng and coworkers realized and identified this Mo$_5$Te$_8$ monolayer a coloring-triangle (CT) lattice[24,30] by either theory or experiments. They discovered that this structure possesses electronic band structures similar to those of a kagome lattice, located near the Fermi level (Figure 10m and 10p). An STM topography image clearly shows the CT lattice, as indicated by the blue solid lines in Figure 10n, corresponding to the atomic structure in Figure 10o. The orange dashed lines connect the shared Te atoms on the MTB, with the Mo atoms at the vertices of the triangular regions forming the lattice points of the CT lattice, and the red arrows highlight the $2\sqrt{3} \times 2\sqrt{3}$ superlattice. Each set of the CT bands (CT1-CT4) consists of a "nominally flat band" with minimal broadening, accompanied by two Dirac bands, with irreducible representations indicating their connection through mirror symmetry operations $\sigma_h$ (Figure 10l). Note that the CT lattice was previously predicted by Liu and coworker in 2019[30] (see Section 2 for more details).

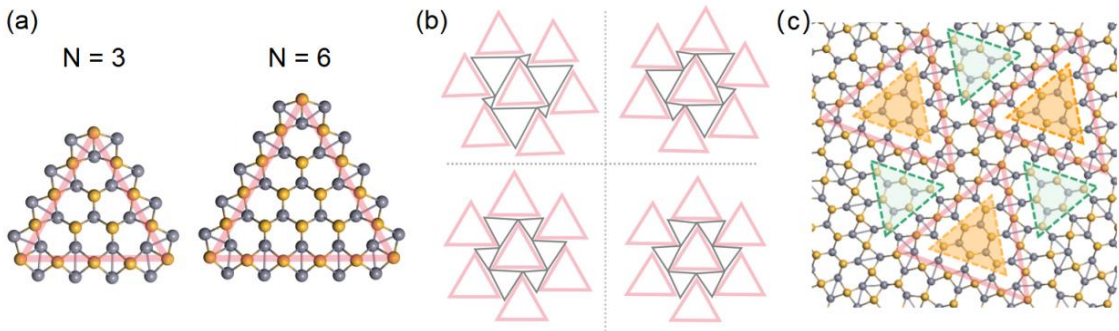

Figure 11: (a) Triangular MTB loops of different sizes in monolayer MoTe$_{2-x}$, with the MTBs

highlighted by pink triangles. (b) Various arrangements of MTB loops, with a grid of four (breathing) coloring triangular lattices highlighted with solid black lines connecting the Mo atoms at the vertices. (c) Atomic structure of $Mo_{33}Te_{56}$, with orange and green triangles marking the regions within and between the MTB loops, respectively.

Very recently, Ji, Zhang and coworkers[204] proved, with both theory and experiments, that the triangular MTB loops can be aligned in uniformly sized and well-ordered array with different loop sizes and arrangements. They regarded differently sized MTB loops as superatoms and used them as building blocks to construct kagome and its variant lattices depending on the size and arrangement of MTB loops. For instance, Fig. 11a shows the atomic structures of $MoTe_2$ MTB loops of different sizes, in which the shared Te atoms are highlighted by pink shadowed triangles. These building blocks can theoretically be arranged into various MTB superlattices (Figure 11b), among them $Mo_5Te_8$ is the smallest superlattice in size. The authors investigate various band fillings and predict several possible electronic phases, including magnetic states, correlated insulators, and topological insulators. They present the formation energies of different configurations and identify four stable monolayer structures as the chemical potential of Te decreases (from right to left), with the calculations performed without considering DFT+$U$. These four structures correspond to varying Te chemical potentials, with lattice constants ranging from 12.8 to 25.9 Å. Experiments have observed only these specific structures, confirming the effectiveness of using varying chemical potentials to determine stable configurations without $U$. Therefore, this finding highlights the importance of accurately matching the chemical potential in theoretical calculations with experimental growth conditions. Zhang, Wu, Yuan, Ji and coworkers investigated the electronic and magnetic properties of $Mo_{33}Te_{56}$[205] (Figure 11c). The non-magnetic band structure of $Mo_{33}Te_{56}$ indicates, at least, three kagome band crossing the Fermi level and are thus partially filled, leading to a high density of states at the Fermi level. Such high DOS drives the monolayer to exhibit spontaneous magnetization and an unidentified correlated insulating state.

These MTBs are of particular promising for constructing neat kagome bands near the Fermi level, as the MTB introduces in-gap electronic states near the Fermi level within a sufficiently large bandgap of the parent TMD (here, $MoTe_2$). Thus, the kagome bands resulted from interactions of these stats should be essentially neat, namely close to the Fermi level and within a well-defined bandgap. The variation in the size and arrangement of the MTB loops offers a rich diversity in electronic structures and thus physical properties of the resulting superlattices. Experimentally, these electronic, magnetic, and topological properties of the $MoTe_{2-x}$ monolayer can, in principle, be further tuned through gating, doping or other methods, providing a playground for exploring novel states in kagome monolayers. Compared to moiré systems, MTB superlattices have smaller periodicities and form isolated kagome electronic bands, with carrier density being more strongly influenced by the internal structure and the arrangement of MTB loops.

# 7. "1+3" strategy: constructing a kagome lattice from a triangular lattice

Triangular lattices are widely presented in two-dimensional materials or on solid surfaces. This section introduces a "1+3" strategy for constructing a kagome lattice from a triangular lattice. As shown in Figure 12a, if we differentiate one lattice site (the "1", in blue) from the rest three ones (the "3", in pink) in a 2×2 supercell of a triangular lattice (gray dashed lines), these three pink sites form a kagome lattice. The differentiation could be done by means of, such as, formation of vacancies, atomic substitutions or atomic/molecular adsorption. Alternatively, if we consider a triangular lattice formed by those blue sites, those pink sites represent midpoints of the lines connecting the nearest neighbor lattice sites. Thus, a kagome lattice is formed by these pink sites. The second interpretation was already demonstrated in kagome lattices formed with Fe atoms (Figure 3d) and the potential wells shown in Figure 7. Three examples are detailly discussed in the following paragraphs to more clearly explain this "1+3" strategy.

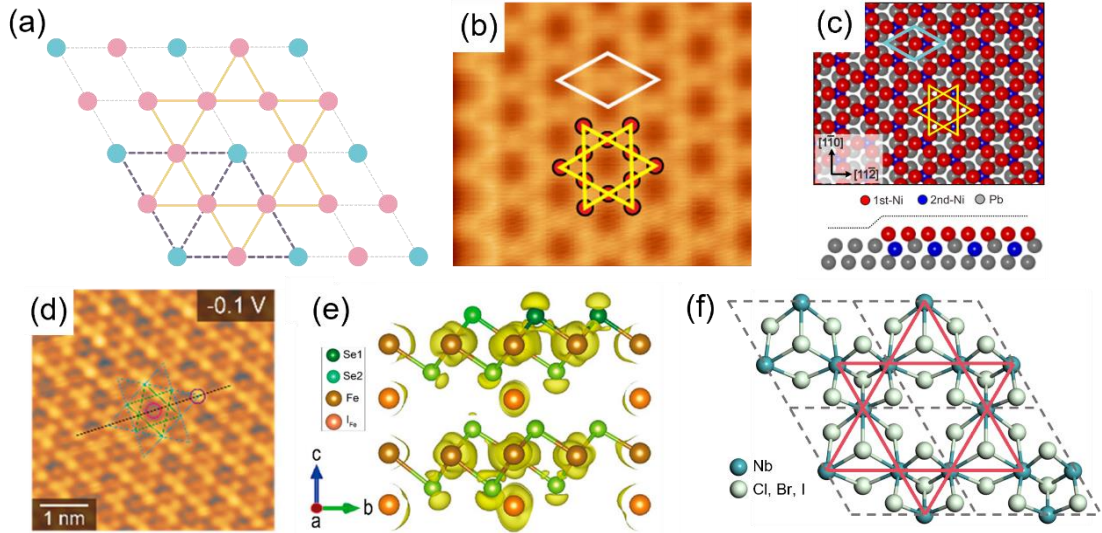

Figure 12 (a) Schematic illustration of constructing a kagome lattice from a triangular lattice. The gray dashed lines represent a 1×1 triangular lattice, the purple dashed lines represent a 2×2 supercell, and the red and blue dots represent different lattices points. The orange solid lines highlight the kagome lattice. (b-c) STM image (b) and atomic structure (c) of Ni deposited on a Pb(111) surface. In (b), $U_b$ = +10 mV, $I_t$ = 1.0 nA. The red dots indicate the kagome lattice formed by the Ni (yellow hexagrams). The white (b) and light blue (c) rhombus outlines the unit cell of the kagome lattice. (d-e) STM image (d) and atomic structure (e) of $Fe_5Se_8$. In (d), $U$ = -0.1V, $I_t$ = 100 pA, and scan area is 5 × 5 nm$^2$. The green dashed lines indicate the two periodic kagome lattices observed in the STM image, corresponding to 2×2 and $2\sqrt{3}\times2\sqrt{3}$ structures. In (e), the integrated charge density from -0.5 to -0.3 eV is overlaid on the atomic structure. (f) Atomic structure of $Nb_3X_8$. (b-c) Reproduced from Ref. [206], (d-e) Reproduced from Ref. [207].

The first example represents the construction of kagome monolayers on solid surfaces using a "1+3" template. Hsu, Bihlmayer and coworkers[206] deposited Ni atoms on a Pb(111) substrate and observed a kagome lattice in the STM image shown in Figure 12b. The line profile extracted from the STM images indicated that the average apparent height of the adatom layer is unusually low, only 0.82 Å. Theoretical calculations revealed that during deposition, one of the four Pb atoms in each 2×2 Pb(111) surface supercell is substituted by a Ni atom, resulting in a "1+3" difference on the triangularly arranged surface atoms (Fig. 12c). Subsequently deposited Ni atoms prefer to adsorb at the hollow sites of the Ni-Pb2 triangles (Figure 12c), forming a kagome monolayer of Ni adatoms on the surface.

The second example focuses on the preparation of kagome few-layers through atomic intercalation, a technique commonly used to modify the properties of 2D materials. In this approach, either three or one intercalating atoms are periodically positioned within the vdW gap of each 2×2 supercell of a triangular latticed 2D bilayer, resulting in a 3/4 ML or 1/4 ML intercalated bilayer. The 3/4 ML intercalating atoms themselves form a kagome monolayer, while the 1/4 ML atoms distinct the four identical unit cells in each 2×2 supercell, dividing them into two groups, one with three unit cells and the other with one. The 3-unit-cell group thus has the potential to form a kagome lattice. Figure 12d shows an example of this approach realized by Zhang, Fu, Liu and coworkers in triangularly latticed 1T-$FeSe_2$ layers [207]. As shown in Figure 12e, multilayer $FeSe_2$ is AA stacked, and after intercalating 1/4 ML of Fe atoms directly below the Fe atoms in the $FeSe_2$ layer, a 2×2 supercell is formed, resulting in a bulk chemical stoichiometry of $Fe_5Se_8$. The intercalated Fe atoms distinct and then divided the surface Se atoms into two types: 3/4 of the Se atoms (Se1) are positioned 0.025 Å higher than the remaining 1/4 of Se atoms (Se2). As we discussed, the Se1 atoms form a kagome lattice, and the corresponding STM image (shown in Fig. 12d) displays a clear kagome feature. Besides, DFT calculations reveal that a similar height difference occurs among the Fe atoms in the $FeSe_2$ layer, causing 3/4 of the Fe atoms to form a kagome lattice and generate kagome bands. This approach was employed again in a 1T-$CoTe_2$ bilayer and multilayer, with 1/4 ML self-intercalation resulting in a kagome lattice similar to that of $Fe_5Se_8$[208].

The final examples concentrate on directly modifying 2D layer themselves, an approach potentially applicable to monolayers. The T or H phase $MX_2$-type 2D materials (where M is a metal and X is a group-16 or -17 nonmetal) generally from triangular lattices. From a materials perspective, several methods can introduce the "1+3" differentiation within a $MX_2$ layer beyond the intercalation-based formation of $M_5X_8$. For instance, the periodically removal of 1/4 ML metal atoms from the $MX_2$ layer leads to the formation of a $M_3X_8$ layer, leaving the rest M atoms arranged in a kagome monolayer. Examples of such structures include $Nb_3X_8$ (X=Cl, Br, I)[209] (Figure 12f) and their variants $Nb_3TeCl_7$[210] and $Nb_3SeI_7$[21]. Analogously removing 1/4 ML of X atoms in a X sublayer results in the formation of $M_4X_7$ monolayer, where the rest 3/4 X atoms construct a kagome sub-monolayer, as experimentally demonstrated in $Pt_4Te_7$[211,212]. For theory, high-throughput calculations comprehensively explored these methods for monolayer metal oxides[213].

In addition to the above examples, kagome lattices can also be constructed through methods such as adsorption[214] and vacancy introduction[215–218], and among others, which are not detailed here. The "1+3" strategy is a general approach for constructing kagome lattices from triangular lattices. Exploring new methods and materials for utilizing this this strategy remains a promising and worthwhile area of future research.

# 8. Outlook

In recent years, research on 2D kagome lattices has rapidly advanced, providing a rich variety of materials for physics research. However, we believe this is merely the beginning, as the future holds vast opportunities and challenges. Reflecting on our original goals in studying 2D kagome materials and looking toward the future, we identify three research directions that are crucial for this field.

First, although two-dimensional atomic or electronic kagome lattices have been realized in various systems, materials with their kagome bands observable and tunable near the Fermi level remain scarce. The key to addressing this challenge lies in finding new construction methods and material systems. For example, one approach could be building kagome lattices using the concept of superatoms, employing new techniques such as annealing to achieve "1+3" differentiation, or utilizing multilayer twist angles. Additionally, the rapid development of artificial intelligence (AI) offers new pathways for discovering these materials and methods. We might develop new design methods to establish correlations between elements, structures, and electronic bands, enabling AI to design ideal, experimentally feasible materials.

Second, synthesis, manipulation, and characterization techniques need further advancement. Interacting phenomena widely explored in bulk kagome materials, such as superconductivity and charge density waves, remain underexplored in 2D kagome

few- or mono-layers. Although many potential systems and methods for constructing 2D kagome lattices have been theoretically predicted[23,219-224], the feasibility of these predictions still await experimental validation, highlighting the urgent need for experimental methods to more feasibly and reliably verify theoretical predictions. Moreover, existing measurement techniques also require improvement—for instance, determining whether the observed peaks in STS belong to kagome flat bands and accurately identifying the topological nature of bands using ARPES. Furthermore, the traditional approaches, manipulation methods, and characterization techniques used in bulk materials research need to be adapted and refined for 2D kagome systems.

Finally, the study of spin frustration effects in kagome magnets has long been a focal point in condensed matter physics[225–227]. The relatively flexible construction methods in 2D systems offer unique opportunities to realize neat 2D kagome magnets[228] and precisely tune key parameters such as spin exchange constants. This provides a promising avenue for exploring and verifying novel phenomena such as quantum spin liquids, making it an attractive and potential-rich research direction.

In summary, two-dimensional kagome lattice materials exhibit immense research potential. As understanding of these materials deepens, we anticipate that this field will be attracting increasing attention and lead to more groundbreaking discoveries. This review not only summarizes the current state of research on 2D kagome lattices, but also as extends an invitation to the broader scientific community to engage in the "quantum adventure" within 2D kagome lattices, urging more researchers to join this exciting exploration and contribute to its advancement.

**Acknowledgements**

We thank Prof. Yi Du at Beihang University, Shaowei Li at University of California, San Diego, Feng Liu at University of Utah, Chuanxu Ma at University of Science and Technology of China, Prof. Minghu Pan at Shaanxi Normal University, Congjun Wu at Westlake University, Tiantian Zhang at Institute of Theoretical Physics, Haihui Wang at The Chinese University of Hong Kong and Yunlong Wang at Renmin University of China for valuable discussions. We gratefully acknowledge the financial support from the Ministry of Science and Technology (MOST) of China (Grant No. 2023YFA1406500), the National Natural Science Foundation of China (Grants No. 11974422 and 12104504), the Fundamental Research Funds for the Central Universities, and the Research Funds of Renmin University of China [Grants No. 22XNKJ30 and 24XNKJ17]. J.D. was supported by the Outstanding Innovative Talents Cultivation Funded Programs 2023 of Renmin University of China.

# References

[1]	Q. Wang, H. Lei, Y. Qi, C. Felser, *Acc. Mater. Res.* **2024**, *5*, 786.
[2]	Y. Wang, H. Wu, G. T. McCandless, J. Y. Chan, M. N. Ali, *Nat Rev Phys* **2023**, *5*, 635.
[3]	J.-X. Yin, B. Lian, M. Z. Hasan, *Nature* **2022**, *612*, 647.
[4]	Y. Hu, X. Wu, B. R. Ortiz, S. Ju, X. Han, J. Ma, N. C. Plumb, M. Radovic, R. Thomale, S. D. Wilson, A. P. Schnyder, M. Shi, *Nat Commun* **2022**, *13*, 2220.
[5]	C. C. Zhu, X. F. Yang, W. Xia, Q. W. Yin, L. S. Wang, C. C. Zhao, D. Z. Dai, C. P. Tu, B. Q. Song, Z. C. Tao, Z. J. Tu, C. S. Gong, H. C. Lei, Y. F. Guo, S. Y. Li, *Phys. Rev. B* **2022**, *105*, 094507.
[6]	B. R. Ortiz, S. M. L. Teicher, Y. Hu, J. L. Zuo, P. M. Sarte, E. C. Schueller, A. M. M. Abeykoon, M. J. Krogstad, S. Rosenkranz, R. Osborn, R. Seshadri, L. Balents, J. He, S. D. Wilson, *Phys. Rev. Lett.* **2020**, *125*, 247002.
[7]	Y. Xu, Z. Ni, Y. Liu, B. R. Ortiz, Q. Deng, S. D. Wilson, B. Yan, L. Balents, L. Wu, *Nat. Phys.* **2022**, *18*, 1470.
[8]	S. Cao, C. Xu, H. Fukui, T. Manjo, Y. Dong, M. Shi, Y. Liu, C. Cao, Y. Song, *Nat Commun* **2023**, *14*, 7671.
[9]	H. Chen, H. Yang, B. Hu, Z. Zhao, J. Yuan, Y. Xing, G. Qian, Z. Huang, G. Li, Y. Ye, S. Ma, S. Ni, H. Zhang, Q. Yin, C. Gong, Z. Tu, H. Lei, H. Tan, S. Zhou, C. Shen, X. Dong, B. Yan, Z. Wang, H.-J. Gao, *Nature* **2021**, *599*, 222.
[10]	N. Morali, R. Batabyal, P. K. Nag, E. Liu, Q. Xu, Y. Sun, B. Yan, C. Felser, N. Avraham, H. Beidenkopf, *Science* **2019**, *365*, 1286.
[11]	D. F. Liu, A. J. Liang, E. K. Liu, Q. N. Xu, Y. W. Li, C. Chen, D. Pei, W. J. Shi, S. K. Mo, P. Dudin, T. Kim, C. Cacho, G. Li, Y. Sun, L. X. Yang, Z. K. Liu, S. S. P. Parkin, C. Felser, Y. L. Chen, *Science* **2019**, *365*, 1282.
[12]	E. Liu, Y. Sun, N. Kumar, L. Muechler, A. Sun, L. Jiao, S.-Y. Yang, D. Liu, A. Liang, Q. Xu, J. Kroder, V. Süß, H. Borrmann, C. Shekhar, Z. Wang, C. Xi, W. Wang, W. Schnelle, S. Wirth, Y. Chen, S. T. B. Goennenwein, C. Felser, *Nature Phys* **2018**, *14*, 1125.
[13]	Q. Wang, Y. Xu, R. Lou, Z. Liu, M. Li, Y. Huang, D. Shen, H. Weng, S. Wang, H. Lei, *Nat Commun* **2018**, *9*, 3681.
[14]	M. Jovanovic, L. M. Schoop, *J. Am. Chem. Soc.* **2022**, *144*, 10978.
[15]	S. Gao, S. Zhang, C. Wang, S. Yan, X. Han, X. Ji, W. Tao, J. Liu, T. Wang, S. Yuan, G. Qu, Z. Chen, Y. Zhang, J. Huang, M. Pan, S. Peng, Y. Hu, H. Li, Y. Huang, H. Zhou, S. Meng, L. Yang, Z. Wang, Y. Yao, Z. Chen, M. Shi, H. Ding, H. Yang, K. Jiang, Y. Li, H. Lei, Y. Shi, H. Weng, T. Qian, *Phys. Rev. X* **2023**, *13*, 041049.
[16]	K. Nakazawa, Y. Kato, Y. Motome, *Phys. Rev. B* **2024**, *110*, 085112.
[17]	Y. Zhang, Y. Gu, H. Weng, K. Jiang, J. Hu, *Phys. Rev. B* **2023**, *107*, 035126.
[18]	S.-W. Kim, H. Oh, E.-G. Moon, Y. Kim, *Nat Commun* **2023**, *14*, 591.
[19]	Y. Fang, X. Feng, D. Wang, Y. Ding, T. Lin, T. Zhai, F. Huang, *Small* **2023**, *19*, 2207934.

[20] J.-P. Wang, Y.-Q. Fang, W. He, Q. Liu, J.-R. Fu, X.-Y. Li, Y. Liu, B. Gao, L. Zhen, C.-Y. Xu, F.-Q. Huang, A. J. Meixner, D. Zhang, Y. Li, *Advanced Optical Materials* **2023**, *11*, 2300031.
[21] J.-P. Wang, X. Chen, Q. Zhao, Y. Fang, Q. Liu, J. Fu, Y. Liu, X. Xu, J. Zhang, L. Zhen, C.-Y. Xu, F. Huang, A. J. Meixner, D. Zhang, G. Gou, Y. Li, *ACS Nano* **2024**, *18*, 16274.
[22] S. Park, S. Kang, H. Kim, K. H. Lee, P. Kim, S. Sim, N. Lee, B. Karuppannan, J. Kim, J. Kim, K. I. Sim, M. J. Coak, Y. Noda, C.-H. Park, J. H. Kim, J.-G. Park, *Sci Rep* **2020**, *10*, 20998.
[23] Y. Chen, S. Xu, Y. Xie, C. Zhong, C. Wu, S. B. Zhang, *Phys. Rev. B* **2018**, *98*, 035135.
[24] L. Lei, J. Dai, H. Dong, Y. Geng, F. Cao, C. Wang, R. Xu, F. Pang, Z.-X. Liu, F. Li, Z. Cheng, G. Wang, W. Ji, *Nat Commun* **2023**, *14*, 6320.
[25] H. Liu, S. Meng, F. Liu, *Phys. Rev. Materials* **2021**, *5*, 084203.
[26] Y. Zhou, G. Sethi, H. Liu, Z. Wang, F. Liu, *Nanotechnology* **2022**, *33*, 415001.
[27] Y. Zhou, G. Sethi, C. Zhang, X. Ni, F. Liu, *Phys. Rev. B* **2020**, *102*, 125115.
[28] W. Jiang, X. Ni, F. Liu, *Acc. Chem. Res.* **2021**, *54*, 416.
[29] Y. Li, S. Zhai, Y. Liu, J. Zhang, Z. Meng, J. Zhuang, H. Feng, X. Xu, W. Hao, M. Zhou, G. Lu, S. X. Dou, Y. Du, *Advanced Science* **2023**, 2303483.
[30] S. Zhang, M. Kang, H. Huang, W. Jiang, X. Ni, L. Kang, S. Zhang, H. Xu, Z. Liu, F. Liu, *Phys. Rev. B* **2019**, *99*, 100404.
[31] S. Yan, D. A. Huse, S. R. White, **2011**, *332*.
[32] Z. Lin, J.-H. Choi, Q. Zhang, W. Qin, S. Yi, P. Wang, L. Li, Y. Wang, H. Zhang, Z. Sun, L. Wei, S. Zhang, T. Guo, Q. Lu, J.-H. Cho, C. Zeng, Z. Zhang, *Phys. Rev. Lett.* **2018**, *121*, 096401.
[33] G. Xu, B. Lian, S.-C. Zhang, *Phys. Rev. Lett.* **2015**, *115*, 186802.
[34] F. H. Yu, D. H. Ma, W. Z. Zhuo, S. Q. Liu, X. K. Wen, B. Lei, J. J. Ying, X. H. Chen, *Nat Commun* **2021**, *12*, 3645.
[35] B. Huang, G. Clark, E. Navarro-Moratalla, D. R. Klein, R. Cheng, K. L. Seyler, D. Zhong, E. Schmidgall, M. A. McGuire, D. H. Cobden, W. Yao, D. Xiao, P. Jarillo-Herrero, X. Xu, *Nature* **2017**, *546*, 270.
[36] G. Sethi, Y. Zhou, L. Zhu, L. Yang, F. Liu, *Phys. Rev. Lett.* **2021**, *126*, 196403.
[37] F. Chen, J. Lu, X. Zhao, G. Hu, X. Yuan, J. Ren, *Applied Physics Letters* **2024**, *125*, 043103.
[38] R. Yin, X. Zhu, Q. Fu, T. Hu, L. Wan, Y. Wu, Y. Liang, Z. Wang, Z.-L. Qiu, Y.-Z. Tan, C. Ma, S. Tan, W. Hu, B. Li, Z. F. Wang, J. Yang, B. Wang, *Nat Commun* **2024**, *15*, 2969.
[39] X. Li, D. Wang, H. Hu, Y. Pan, *Nanotechnology* **2024**, *35*, 145601.
[40] M. Telychko, G. Li, P. Mutombo, D. Soler-Polo, X. Peng, J. Su, S. Song, M. J. Koh, M. Edmonds, P. Jelínek, J. Wu, J. Lu, *Sci. Adv.* **2021**, *7*, eabf0269.
[41] S. Okamoto, N. Mohanta, E. Dagotto, D. N. Sheng, *Commun Phys* **2022**, *5*, 1.

[42]     C. Wu, D. Bergman, L. Balents, S. Das Sarma, *Phys. Rev. Lett.* **2007**, *99*, 070401.
[43]     G.-F. Zhang, Y. Li, C. Wu, *PHYSICAL REVIEW B* **2014**.
[44]     S. Zhang, H. Hung, C. Wu, *Phys. Rev. A* **2010**, *82*, 053618.
[45]     W.-C. Lee, C. Wu, S. Das Sarma, *Phys. Rev. A* **2010**, *82*, 053611.
[46]     C. Wu, *Phys. Rev. Lett.* **2008**, *101*, 186807.
[47]     J. Mao, H. Zhang, Y. Jiang, Y. Pan, M. Gao, W. Xiao, H.-J. Gao, *J. Am. Chem. Soc.* **2009**, *131*, 14136.
[48]     S. D. Feyter, F. C. D. Schryver, *Chem. Soc. Rev.* **2003**, *32*, 139.
[49]     X. Liu, C. Guan, D. Wang, L. Wan, *Advanced Materials* **2014**, *26*, 6912.
[50]     R. Gutzler, *Phys. Chem. Chem. Phys.* **2016**, *18*, 29092.
[51]     S. Furukawa, H. Uji-i, K. Tahara, T. Ichikawa, M. Sonoda, F. C. De Schryver, Y. Tobe, S. De Feyter, *J. Am. Chem. Soc.* **2006**, *128*, 3502.
[52]     K. Tahara, S. Furukawa, H. Uji-i, T. Uchino, T. Ichikawa, J. Zhang, W. Mamdouh, M. Sonoda, F. C. De Schryver, S. De Feyter, Y. Tobe, *J. Am. Chem. Soc.* **2006**, *128*, 16613.
[53]     X. Huang, S. Zhang, L. Liu, L. Yu, G. Chen, W. Xu, D. Zhu, *Angewandte Chemie International Edition* **2018**, *57*, 146.
[54]     L. Dong, Y. Kim, D. Er, A. M. Rappe, V. B. Shenoy, *Phys. Rev. Lett.* **2016**, *116*, 096601.
[55]     T. Deng, W. Shi, Z. M. Wong, G. Wu, X. Yang, J.-C. Zheng, H. Pan, S.-W. Yang, *J. Phys. Chem. Lett.* **2021**, *12*, 6934.
[56]     Y. Yin, Y. Gao, L. Zhang, Y.-Y. Zhang, S. Du, *Sci. China Mater.* **2024**, *67*, 1202.
[57]     M. G. Yamada, H. Fujita, M. Oshikawa, *Phys. Rev. Lett.* **2017**, *119*, 057202.
[58]     Y.-P. Mo, X.-H. Liu, D. Wang, *ACS Nano* **2017**, *11*, 11694.
[59]     H. Zhou, H. Dang, J.-H. Yi, A. Nanci, A. Rochefort, J. D. Wuest, *J. Am. Chem. Soc.* **2007**, *129*, 13774.
[60]     F. Haase, B. V. Lotsch, *Chem. Soc. Rev.* **2020**, *49*, 8469.
[61]     J. Tu, W. Song, B. Chen, Y. Li, L. Chen, *Chemistry A European J* **2023**, *29*, e202302380.
[62]     U. Schlickum, R. Decker, F. Klappenberger, G. Zoppellaro, S. Klyatskaya, W. Auwärter, S. Neppl, K. Kern, H. Brune, M. Ruben, J. V. Barth, *J. Am. Chem. Soc.* **2008**, *130*, 11778.
[63]     C. Deng, J. Wang, H. Zhu, C. Xu, X. Fan, Y. Wen, P. Huang, H. Lin, Q. Li, L. Chi, *J. Phys. Chem. Lett.* **2023**, *14*, 9584.
[64]     T. Chen, Q. Chen, X. Zhang, D. Wang, L.-J. Wan, *J. Am. Chem. Soc.* **2010**, *132*, 5598.
[65]     M. Dong, X. Miao, R. Brisse, W. Deng, B. Jousselme, F. Silly, *NPG Asia Mater* **2020**, *12*, 1.
[66]     T. Wang, Q. Fan, L. Feng, Z. Tao, J. Huang, H. Ju, Q. Xu, S. Hu, J. Zhu, *ChemPhysChem* **2017**, *18*, 3329.
[67]     L. Liu, W. Xiao, J. Mao, H. Zhang, Y. Jiang, H. Zhou, K. Yang, H. Gao, *Chinese Chemical Letters* **2018**, *29*, 183.

[68]     Y. Wei, J. E. Reutt-Robey, *J. Am. Chem. Soc.* **2011**, *133*, 15232.
[69]     H. Zhang, W. D. Xiao, J. Mao, H. Zhou, G. Li, Y. Zhang, L. Liu, S. Du, H.-J. Gao, *J. Phys. Chem. C* **2012**, *116*, 11091.
[70]     W. Pan, C. Mützel, S. Haldar, H. Hohmann, S. Heinze, J. M. Farrell, R. Thomale, M. Bode, F. Würthner, J. Qi, *Angew Chem Int Ed* **2024**, *63*, e202400313.
[71]     M. Pan, X. Zhang, Y. Zhou, P. Wang, Q. Bian, H. Liu, X. Wang, X. Li, A. Chen, X. Lei, S. Li, Z. Cheng, Z. Shao, H. Ding, J. Gao, F. Li, F. Liu, *Phys. Rev. Lett.* **2023**, *130*, 036203.
[72]     K. I. Shivakumar, S. Noro, Y. Yamaguchi, Y. Ishigaki, A. Saeki, K. Takahashi, T. Nakamura, I. Hisaki, *Chemical Communications* **2021**, *57*, 1157.
[73]     P. Tholen, C. A. Peeples, R. Schaper, C. Bayraktar, T. S. Erkal, M. M. Ayhan, B. Çoşut, J. Beckmann, A. O. Yazaydın, M. Wark, G. Hanna, Y. Zorlu, G. Yücesan, *Nat Commun* **2020**, *11*, 3180.
[74]     C.-H. Liu, A. Wei, M. F. Cheung, D. F. Perepichka, *Chem. Mater.* **2022**, *34*, 3461.
[75]     Y. Jing, T. Heine, *J. Am. Chem. Soc.* **2019**, *141*, 743.
[76]     T. Hu, T. Zhang, H. Mu, Z. Wang, *J. Phys. Chem. Lett.* **2022**, *13*, 10905.
[77]     G. Galeotti, F. De Marchi, E. Hamzehpoor, O. MacLean, M. Rajeswara Rao, Y. Chen, L. V. Besteiro, D. Dettmann, L. Ferrari, F. Frezza, P. M. Sheverdyaeva, R. Liu, A. K. Kundu, P. Moras, M. Ebrahimi, M. C. Gallagher, F. Rosei, D. F. Perepichka, G. Contini, *Nat. Mater.* **2020**, *19*, 874.
[78]     Z. Shi, N. Lin, *J. Am. Chem. Soc.* **2009**, *131*, 5376.
[79]     M. Hua, B. Xia, M. Wang, E. Li, J. Liu, T. Wu, Y. Wang, R. Li, H. Ding, J. Hu, Y. Wang, J. Zhu, H. Xu, W. Zhao, N. Lin, *J. Phys. Chem. Lett.* **2021**, *12*, 3733.
[80]     L. Z. Zhang, Z. F. Wang, B. Huang, B. Cui, Z. Wang, S. X. Du, H.-J. Gao, F. Liu, *Nano Lett.* **2016**, *16*, 2072.
[81]     J. Wang, Y. Zheng, X. Nie, C. Xu, Z. Hao, L. Song, S. You, J. Xi, M. Pan, H. Lin, Y. Li, H. Zhang, Q. Li, L. Chi, *J. Phys. Chem. Lett.* **2021**, *12*, 8151.
[82]     L. Dong, Z. Gao, N. Lin, *Progress in Surface Science* **2016**, *91*, 101.
[83]     J. Li, A. Kumar, B. A. Johnson, S. Ott, *Nat Commun* **2023**, *14*, 1.
[84]     B. Liu, G. Miao, W. Zhong, X. Huang, N. Su, J. Guo, W. Wang, *ACS Nano* **2022**, *16*, 2147.
[85]     L. She, Z. Shen, Z. Xie, L. Wang, Y. Song, X.-S. Wang, Y. Jia, Z. Zhang, W. Zhang, *Phys. Rev. Lett.* **2022**, *129*, 026802.
[86]     K. Wada, K. Sakaushi, S. Sasaki, H. Nishihara, *Angewandte Chemie International Edition* **2018**, *57*, 8886.
[87]     W. P. Lustig, S. Mukherjee, N. D. Rudd, A. V. Desai, J. Li, S. K. Ghosh, *Chem. Soc. Rev.* **2017**, *46*, 3242.
[88]     T. Takenaka, K. Ishihara, M. Roppongi, Y. Miao, Y. Mizukami, T. Makita, J. Tsurumi, S. Watanabe, J. Takeya, M. Yamashita, K. Torizuka, Y. Uwatoko, T. Sasaki, X. Huang, W. Xu, D. Zhu, N. Su, J.-G. Cheng, T. Shibauchi, K. Hashimoto, *Sci. Adv.* **2021**, *7*, eabf3996.
[89]     U. Ryu, S. Jee, P. C. Rao, J. Shin, C. Ko, M. Yoon, K. S. Park, K. M. Choi, *Coordination Chemistry Reviews* **2021**, *426*, 213544.

[90]     M. G. Campbell, D. Sheberla, S. F. Liu, T. M. Swager, M. Dincă, *Angewandte Chemie International Edition* **2015**, *54*, 4349.
[91]     X. Huang, P. Sheng, Z. Tu, F. Zhang, J. Wang, H. Geng, Y. Zou, C. Di, Y. Yi, Y. Sun, W. Xu, D. Zhu, *Nat Commun* **2015**, *6*, 7408.
[92]     J. J. Richardson, M. Björnmalm, F. Caruso, *Science* **2015**, *348*, aaa2491.
[93]     M. C. So, S. Jin, H.-J. Son, G. P. Wiederrecht, O. K. Farha, J. T. Hupp, *J. Am. Chem. Soc.* **2013**, *135*, 15698.
[94]     O. Shekhah, H. Wang, S. Kowarik, F. Schreiber, M. Paulus, M. Tolan, C. Sternemann, F. Evers, D. Zacher, R. A. Fischer, C. Wöll, *J. Am. Chem. Soc.* **2007**, *129*, 15118.
[95]     J. Liu, C. Wöll, *Chem. Soc. Rev.* **2017**, *46*, 5730.
[96]     T. Kambe, R. Sakamoto, K. Hoshiko, K. Takada, M. Miyachi, J.-H. Ryu, S. Sasaki, J. Kim, K. Nakazato, M. Takata, H. Nishihara, *J. Am. Chem. Soc.* **2013**, *135*, 2462.
[97]     Z. F. Wang, N. Su, F. Liu, *Nano Lett.* **2013**, *13*, 2842.
[98]     Z. Fu, Y. Zhang, M. Jia, S. Zhang, L. Guan, D. Xing, J. Tao, *Phys. Chem. Chem. Phys.* **2024**, *26*, 21767.
[99]     E. Coronado, *Nat Rev Mater* **2020**, *5*, 87.
[100]    B. Field, A. Schiffrin, N. V. Medhekar, *npj Comput Mater* **2022**, *8*, 227.
[101]    Q. Yu, D. Wang, *J. Mater. Chem. A* **2023**, *11*, 5548.
[102]    D. Kumar, J. Hellerstedt, B. Field, B. Lowe, Y. Yin, N. V. Medhekar, A. Schiffrin, *Adv Funct Materials* **2021**, *31*, 2106474.
[103]    N. Su, W. Jiang, Z. Wang, F. Liu, *Applied Physics Letters* **2018**, *112*, 033301.
[104]    L. Liu, B. Zhao, J. Zhang, H. Bao, H. Huan, Y. Xue, Y. Li, Z. Yang, *Phys. Rev. B* **2021**, *104*, 245414.
[105]    X. Zhang, Y. Zhou, B. Cui, M. Zhao, F. Liu, *Nano Lett.* **2017**, *17*, 6166.
[106]    Y. Zhang, J. Lu, W. Gao, Y. Zhang, N. Li, S. Li, G. Niu, B. Fu, L. Gao, J. Cai, *Chin. J. Chem.* **2024**, cjoc.202400557.
[107]    T. Kambe, R. Sakamoto, T. Kusamoto, T. Pal, N. Fukui, K. Hoshiko, T. Shimojima, Z. Wang, T. Hirahara, K. Ishizaka, S. Hasegawa, F. Liu, H. Nishihara, *J. Am. Chem. Soc.* **2014**, *136*, 14357.
[108]    Y. Miyake, T. Nagata, H. Tanaka, M. Yamazaki, M. Ohta, R. Kokawa, T. Ogawa, *ACS Nano* **2012**, *6*, 3876.
[109]    X. Peng, Y. Geng, M. Zhang, F. Cheng, L. Cheng, K. Deng, Q. Zeng, *Nano Res.* **2019**, *12*, 537.
[110]    I. Piquero-Zulaica, W. Hu, A. P. Seitsonen, F. Haag, J. Küchle, F. Allegretti, Y. Lyu, L. Chen, K. Wu, Z. M. A. El-Fattah, E. Aktürk, S. Klyatskaya, M. Ruben, M. Muntwiler, J. V. Barth, Y. Zhang, *Advanced Materials* **2024**, 2405178.
[111]    C. Lyu, Y. Gao, K. Zhou, M. Hua, Z. Shi, P.-N. Liu, L. Huang, N. Lin, *ACS Nano* **2024**, *18*, 19793.
[112]    B. Field, *npj Computational Materials* **2022**.
[113]    L. Yan, O. J. Silveira, B. Alldritt, S. Kezilebieke, A. S. Foster, P. Liljeroth, *ACS Nano* **2021**, *15*, 17813.

[114]    B. Lowe, B. Field, J. Hellerstedt, J. Ceddia, H. L. Nourse, B. J. Powell, N. V. Medhekar, A. Schiffrin, *Nat Commun* **2024**, *15*, 3559.
[115]    M. Zhang, Z. Wang, X. Bo, R. Huang, D. Deng, *Angewandte Chemie International Edition* **2024**, *n/a*, e202419661.
[116]    D. Deng, K. S. Novoselov, Q. Fu, N. Zheng, Z. Tian, X. Bao, *Nature Nanotech* **2016**, *11*, 218.
[117]    Y. Yang, B. Liang, J. Kreie, M. Hambsch, Z. Liang, C. Wang, S. Huang, X. Dong, L. Gong, C. Liang, D. Lou, Z. Zhou, J. Lu, Y. Yang, X. Zhuang, H. Qi, U. Kaiser, S. C. B. Mannsfeld, W. Liu, A. Gölzhäuser, Z. Zheng, *Nature* **2024**, *630*, 878.
[118]    K. Liu, H. Qi, R. Dong, R. Shivhare, M. Addicoat, T. Zhang, H. Sahabudeen, T. Heine, S. Mannsfeld, U. Kaiser, Z. Zheng, X. Feng, *Nat. Chem.* **2019**, *11*, 994.
[119]    D. M. Eigler, E. K. Schweizer, *Nature* **1990**, *344*, 524.
[120]    M. F. Crommie, C. P. Lutz, D. M. Eigler, *Science* **1993**, *262*, 218.
[121]    I. Piquero-Zulaica, J. Lobo-Checa, Z. M. A. El-Fattah, J. E. Ortega, F. Klappenberger, W. Auwärter, J. V. Barth, *Rev. Mod. Phys.* **2022**, *94*, 045008.
[122]    S. Sun, S. Zhao, Y. Z. Luo, X. Gu, X. Lian, A. Tadich, D.-C. Qi, Z. Ma, Y. Zheng, C. Gu, J. L. Zhang, Z. Li, W. Chen, *Nano Lett.* **2020**, *20*, 5583.
[123]    J. L. Zhang, S. Zhao, S. Sun, H. Ding, J. Hu, Y. Li, Q. Xu, X. Yu, M. Telychko, J. Su, C. Gu, Y. Zheng, X. Lian, Z. Ma, R. Guo, J. Lu, Z. Sun, J. Zhu, Z. Li, W. Chen, *ACS Nano* **2020**, *14*, 3687.
[124]    H. Tian, J.-Q. Zhang, W. Ho, J.-P. Xu, B. Xia, Y. Xia, J. Fan, H. Xu, M. Xie, S. Y. Tong, *Matter* **2020**, *2*, 111.
[125]    K. K. Gomes, W. Mar, W. Ko, F. Guinea, H. C. Manoharan, *Nature* **2012**, *483*, 306.
[126]    S. Wang, L. Z. Tan, W. Wang, S. G. Louie, N. Lin, *Phys. Rev. Lett.* **2014**, *113*, 196803.
[127]    M. R. Slot, T. S. Gardenier, P. H. Jacobse, *NATURE PHYSICS* **2017**, *13*.
[128]    S. N. Kempkes, M. R. Slot, J. J. van den Broeke, P. Capiod, W. A. Benalcazar, D. Vanmaekelbergh, D. Bercioux, I. Swart, C. Morais Smith, *Nat. Mater.* **2019**, *18*, 1292.
[129]    F. E. Kalff, M. P. Rebergen, E. Fahrenfort, J. Girovsky, R. Toskovic, J. L. Lado, J. Fernández-Rossier, A. F. Otte, *Nature Nanotech* **2016**, *11*, 926.
[130]    S. Fölsch, J. Yang, C. Nacci, K. Kanisawa, *Phys. Rev. Lett.* **2009**, *103*, 096104.
[131]    K. Sagisaka, D. Fujita, *Applied Physics Letters* **2006**, *88*, 203118.
[132]    V. D. Pham, Y. Pan, S. C. Erwin, S. Fölsch, *Phys. Rev. Research* **2024**, *6*, 013269.
[133]    A. A. Khajetoorians, *NATURE PHYSICS* **2012**, *8*.
[134]    C. F. Hirjibehedin, C. P. Lutz, A. J. Heinrich, *Science* **2006**, *312*, 1021.
[135]    Q. Tian, S. Izadi Vishkayi, M. Bagheri Tagani, L. Zhang, Y. Tian, L.-J. Yin, L. Zhang, Z. Qin, *Nano Lett.* **2023**, *23*, 9851.
[136]    L. Huang, X. Kong, Q. Zheng, Y. Xing, H. Chen, Y. Li, Z. Hu, S. Zhu, J. Qiao, Y.-Y. Zhang, H. Cheng, Z. Cheng, X. Qiu, E. Liu, H. Lei, X. Lin, Z. Wang, H. Yang, W. Ji, H.-J. Gao, *Nat Commun* **2023**, *14*, 5230.

[137]   B. He, G. Tian, J. Gou, B. Liu, K. Shen, Q. Tian, Z. Yu, F. Song, H. Xie, Y. Gao, Y. Lu, K. Wu, L. Chen, H. Huang, *Surface Science* **2019**, *679*, 147.
[138]   L. Zhou, F. Yang, S. Zhang, T. Zhang, *Advanced Materials* **2024**, *36*, 2309803.
[139]   T. Li, S. Jiang, B. Shen, Y. Zhang, L. Li, Z. Tao, T. Devakul, K. Watanabe, T. Taniguchi, L. Fu, J. Shan, K. F. Mak, *Nature* **2021**, *600*, 641.
[140]   Y. Xu, S. Liu, D. A. Rhodes, K. Watanabe, T. Taniguchi, J. Hone, V. Elser, K. F. Mak, J. Shan, *Nature* **2020**, *587*, 214.
[141]   Y. Cao, V. Fatemi, S. Fang, K. Watanabe, T. Taniguchi, E. Kaxiras, P. Jarillo-Herrero, *Nature* **2018**, *556*, 43.
[142]   R. Bistritzer, A. H. MacDonald, *Proceedings of the National Academy of Sciences* **2011**, *108*, 12233.
[143]   K. P. Nuckolls, A. Yazdani, *Nat Rev Mater* **2024**, *9*, 460.
[144]   S. K. Behura, A. Miranda, S. Nayak, K. Johnson, P. Das, N. R. Pradhan, *emergent mater.* **2021**, *4*, 813.
[145]   Li T.-X., Key Laboratory of Artificial Structures and Quantum Control (Ministry of Education), Shenyang National Laboratory for Materials Science, School of Physics & Astronomy, Shanghai Jiao Tong University, Shanghai 200240, China, Tsung-Dao Lee Institute, Shanghai Jiao Tong University, Shanghai 201210, China, *Acta Phys. Sin.* **2022**, *71*, 127309.
[146]   E. C. Regan, D. Wang, C. Jin, M. I. Bakti Utama, B. Gao, X. Wei, S. Zhao, W. Zhao, Z. Zhang, K. Yumigeta, M. Blei, J. D. Carlström, K. Watanabe, T. Taniguchi, S. Tongay, M. Crommie, A. Zettl, F. Wang, *Nature* **2020**, *579*, 359.
[147]   A. Uri, S. Grover, Y. Cao, J. A. Crosse, K. Bagani, D. Rodan-Legrain, Y. Myasoedov, K. Watanabe, T. Taniguchi, P. Moon, M. Koshino, P. Jarillo-Herrero, E. Zeldov, *Nature* **2020**, *581*, 47.
[148]   H. Yoo, R. Engelke, S. Carr, S. Fang, K. Zhang, P. Cazeaux, S. H. Sung, R. Hovden, A. W. Tsen, T. Taniguchi, K. Watanabe, G.-C. Yi, M. Kim, M. Luskin, E. B. Tadmor, E. Kaxiras, P. Kim, *Nat. Mater.* **2019**, *18*, 448.
[149]   M. R. Rosenberger, H.-J. Chuang, M. Phillips, V. P. Oleshko, K. M. McCreary, S. V. Sivaram, C. S. Hellberg, B. T. Jonker, *ACS Nano* **2020**, *14*, 4550.
[150]   F. K. de Vries, J. Zhu, E. Portolés, G. Zheng, M. Masseroni, A. Kurzmann, T. Taniguchi, K. Watanabe, A. H. MacDonald, K. Ensslin, T. Ihn, P. Rickhaus, *Phys. Rev. Lett.* **2020**, *125*, 176801.
[151]   Y. Liu, P. Stradins, S.-H. Wei, *Science Advances* **2016**, *2*, e1600069.
[152]   Z. Bi, N. F. Q. Yuan, L. Fu, *Phys. Rev. B* **2019**, *100*, 035448.
[153]   Y. Cao, V. Fatemi, A. Demir, S. Fang, S. L. Tomarken, J. Y. Luo, J. D. Sanchez-Yamagishi, K. Watanabe, T. Taniguchi, E. Kaxiras, R. C. Ashoori, P. Jarillo-Herrero, *Nature* **2018**, *556*, 80.
[154]   K. Tran, G. Moody, F. Wu, X. Lu, J. Choi, K. Kim, A. Rai, D. A. Sanchez, J. Quan, A. Singh, J. Embley, A. Zepeda, M. Campbell, T. Autry, T. Taniguchi, K. Watanabe, N. Lu, S. K. Banerjee, K. L. Silverman, S. Kim, E. Tutuc, L. Yang, A. H. MacDonald, X. Li, *Nature* **2019**, *567*, 71.

[155]    C. Jin, E. C. Regan, A. Yan, M. Iqbal Bakti Utama, D. Wang, S. Zhao, Y. Qin, S. Yang, Z. Zheng, S. Shi, K. Watanabe, T. Taniguchi, S. Tongay, A. Zettl, F. Wang, *Nature* **2019**, *567*, 76.
[156]    K. L. Seyler, P. Rivera, H. Yu, N. P. Wilson, E. L. Ray, D. G. Mandrus, J. Yan, W. Yao, X. Xu, *Nature* **2019**, *567*, 66.
[157]    C. L. Tschirhart, M. Serlin, H. Polshyn, A. Shragai, Z. Xia, J. Zhu, Y. Zhang, K. Watanabe, T. Taniguchi, M. E. Huber, A. F. Young, *Science* **2021**, *372*, 1323.
[158]    A. L. Sharpe, E. J. Fox, A. W. Barnard, J. Finney, K. Watanabe, T. Taniguchi, M. A. Kastner, D. Goldhaber-Gordon, *Science* **2019**, *365*, 605.
[159]    T. Li, S. Jiang, B. Shen, Y. Zhang, L. Li, Z. Tao, T. Devakul, K. Watanabe, T. Taniguchi, L. Fu, J. Shan, K. F. Mak, *Nature* **2021**, *600*, 641.
[160]    M. Serlin, C. L. Tschirhart, H. Polshyn, Y. Zhang, J. Zhu, K. Watanabe, T. Taniguchi, L. Balents, A. F. Young, *Science* **2020**, *367*, 900.
[161]    Z. Li, J. Zhuang, L. Wang, H. Feng, Q. Gao, X. Xu, W. Hao, X. Wang, C. Zhang, K. Wu, S. X. Dou, L. Chen, Z. Hu, Y. Du, *Sci. Adv.* **2018**, *4*, eaau4511.
[162]    Q. Zheng, C.-Y. Hao, X.-F. Zhou, Y.-X. Zhao, J.-Q. He, L. He, *Phys. Rev. Lett.* **2022**, *129*, 076803.
[163]    D. Pei, B. Wang, Z. Zhou, Z. He, L. An, S. He, C. Chen, Y. Li, L. Wei, A. Liang, J. Avila, P. Dudin, V. Kandyba, A. Giampietri, M. Cattelan, A. Barinov, Z. Liu, J. Liu, H. Weng, N. Wang, J. Xue, Y. Chen, *Phys. Rev. X* **2022**, *12*, 021065.
[164]    D. Park, C. Park, E. Ko, K. Yananose, R. Engelke, X. Zhang, K. Davydov, M. Green, S. H. Park, J. H. Lee, K. Watanabe, S. M. Yang, K. Wang, P. Kim, Y.-W. Son, H. Yoo, *arXiv 2402.15760*.
[165]    Z. Chu, E. C. Regan, X. Ma, D. Wang, Z. Xu, M. I. B. Utama, K. Yumigeta, M. Blei, K. Watanabe, T. Taniguchi, S. Tongay, F. Wang, K. Lai, *Phys. Rev. Lett.* **2020**, *125*, 186803.
[166]    X. Huang, T. Wang, S. Miao, C. Wang, Z. Li, Z. Lian, T. Taniguchi, K. Watanabe, S. Okamoto, D. Xiao, S.-F. Shi, Y.-T. Cui, *Nat. Phys.* **2021**, *17*, 715.
[167]    X. Huang, L. Chen, S. Tang, C. Jiang, C. Chen, H. Wang, Z.-X. Shen, H. Wang, Y.-T. Cui, *Nano Lett.* **2021**, *21*, 4292.
[168]    K. Lee, M. I. B. Utama, S. Kahn, A. Samudrala, N. Leconte, B. Yang, S. Wang, K. Watanabe, T. Taniguchi, M. V. P. Altoé, G. Zhang, A. Weber-Bargioni, M. Crommie, P. D. Ashby, J. Jung, F. Wang, A. Zettl, *Science Advances* **2020**, *6*, eabd1919.
[169]    F. M. Arnold, A. Ghasemifard, A. Kuc, J. Kunstmann, T. Heine, *2D Mater.* **2023**, *10*, 045010.
[170]    M. Claassen, L. Xian, D. M. Kennes, A. Rubio, *Nat Commun* **2022**, *13*, 4915.
[171]    A. P. Reddy, T. Devakul, L. Fu, *Phys. Rev. Lett.* **2023**, *131*, 246501.
[172]    M. Angeli, A. H. MacDonald, *Proceedings of the National Academy of Sciences* **2021**, *118*, e2021826118.
[173]    B. Padhi, C. Setty, P. W. Phillips, *Nano Lett.* **2018**, *18*, 6175.
[174]    H. Li, Z. Xiang, A. P. Reddy, T. Devakul, R. Sailus, R. Banerjee, T. Taniguchi, K. Watanabe, S. Tongay, A. Zettl, L. Fu, M. F. Crommie, F. Wang, *Science* **2024**, *385*, 86.

[175]  Z. Liu, X. Kong, Z. Wu, L. Zhou, J. Qiao, W. Ji, *Exotic electronic states in gradient-strained untwisted graphene bilayers*, arXiv, **2023**.

[176]  C. R. Woods, F. Withers, M. J. Zhu, Y. Cao, G. Yu, A. Kozikov, M. Ben Shalom, S. V. Morozov, M. M. Van Wijk, A. Fasolino, M. I. Katsnelson, K. Watanabe, T. Taniguchi, A. K. Geim, A. Mishchenko, K. S. Novoselov, *Nat Commun* **2016**, *7*, 10800.

[177]  J. D. Sanchez-Yamagishi, T. Taychatanapat, K. Watanabe, T. Taniguchi, A. Yacoby, P. Jarillo-Herrero, *Phys. Rev. Lett.* **2012**, *108*, 076601.

[178]  L. Wang, Y. Gao, B. Wen, Z. Han, T. Taniguchi, K. Watanabe, M. Koshino, J. Hone, C. R. Dean, *Science* **2015**.

[179]  D. Wang, G. Chen, C. Li, M. Cheng, W. Yang, S. Wu, G. Xie, J. Zhang, J. Zhao, X. Lu, P. Chen, G. Wang, J. Meng, J. Tang, R. Yang, C. He, D. Liu, D. Shi, K. Watanabe, T. Taniguchi, J. Feng, Y. Zhang, G. Zhang, *Phys. Rev. Lett.* **2016**, *116*, 126101.

[180]  W. Zhou, X. Zou, S. Najmaei, Z. Liu, Y. Shi, J. Kong, J. Lou, P. M. Ajayan, B. I. Yakobson, J.-C. Idrobo, *Nano Lett.* **2013**, *13*, 2615.

[181]  S. Najmaei, Z. Liu, W. Zhou, X. Zou, G. Shi, S. Lei, B. I. Yakobson, J.-C. Idrobo, P. M. Ajayan, J. Lou, *Nature Mater* **2013**, *12*, 754.

[182]  Y. L. Huang, Y. Chen, W. Zhang, S. Y. Quek, C.-H. Chen, L.-J. Li, W.-T. Hsu, W.-H. Chang, Y. J. Zheng, W. Chen, A. T. S. Wee, *Nat Commun* **2015**, *6*, 6298.

[183]  H.-P. Komsa, J. Kotakoski, S. Kurasch, O. Lehtinen, U. Kaiser, A. V. Krasheninnikov, *Phys. Rev. Lett.* **2012**, *109*, 035503.

[184]  A. M. van der Zande, P. Y. Huang, D. A. Chenet, T. C. Berkelbach, Y. You, G.-H. Lee, T. F. Heinz, D. R. Reichman, D. A. Muller, J. C. Hone, *Nature Mater* **2013**, *12*, 554.

[185]  Y.-C. Lin, T. Björkman, H.-P. Komsa, P.-Y. Teng, C.-H. Yeh, F.-S. Huang, K.-H. Lin, J. Jadczak, Y.-S. Huang, P.-W. Chiu, A. V. Krasheninnikov, K. Suenaga, *Nat Commun* **2015**, *6*, 6736.

[186]  H. Liu, L. Jiao, F. Yang, Y. Cai, X. Wu, W. Ho, C. Gao, J. Jia, N. Wang, H. Fan, W. Yao, M. Xie, *Phys. Rev. Lett.* **2014**, *113*, 066105.

[187]  H. Murata, A. Koma, *Phys. Rev. B* **1999**, *59*, 10327.

[188]  J. Hong, C. Wang, H. Liu, X. Ren, J. Chen, G. Wang, J. Jia, M. Xie, C. Jin, W. Ji, J. Yuan, Z. Zhang, *Nano Lett.* **2017**, *17*, 6653.

[189]  L. Dong, G.-Y. Wang, Z. Zhu, C.-X. Zhao, X.-Y. Yang, A.-M. Li, J.-L. Chen, D.-D. Guan, Y.-Y. Li, H. Zheng, M.-H. Xie, J.-F. Jia, *Chinese Phys. Lett.* **2018**, *35*, 066801.

[190]  J. Zhang, Y. Xia, B. Wang, Y. Jin, H. Tian, W. kin Ho, H. Xu, C. Jin, M. Xie, *2D Mater.* **2020**, *8*, 015006.

[191]  O. Lehtinen, H.-P. Komsa, A. Pulkin, M. B. Whitwick, M.-W. Chen, T. Lehnert, M. J. Mohn, O. V. Yazyev, A. Kis, U. Kaiser, A. V. Krasheninnikov, *ACS Nano* **2015**, *9*, 3274.

[192]  T. Mori, H. Abe, K. S. K. Saiki, A. K. A. Koma, *Jpn. J. Appl. Phys.* **1993**, *32*, 2945.

[193]　F. S. Ohuchi, B. A. Parkinson, K. Ueno, A. Koma, *Journal of Applied Physics* **1990**, *68*, 2168.

[194]　J. Lin, S. T. Pantelides, W. Zhou, *ACS Nano* **2015**, *9*, 5189.

[195]　S. Barja, S. Wickenburg, Z.-F. Liu, Y. Zhang, H. Ryu, M. M. Ugeda, Z. Hussain, Z.-X. Shen, S.-K. Mo, E. Wong, M. B. Salmeron, F. Wang, M. F. Crommie, D. F. Ogletree, J. B. Neaton, A. Weber-Bargioni, *Nature Phys* **2016**, *12*, 751.

[196]　H. C. Diaz, Y. Ma, R. Chaghi, M. Batzill, *Applied Physics Letters* **2016**, *108*, 191606.

[197]　H. Liu, H. Zheng, F. Yang, L. Jiao, J. Chen, W. Ho, C. Gao, J. Jia, M. Xie, *ACS Nano* **2015**, *9*, 6619.

[198]　H. C. Diaz, R. Chaghi, Y. Ma, M. Batzill, *2D Mater.* **2015**, *2*, 044010.

[199]　Y. Ma, S. Kolekar, H. Coy Diaz, J. Aprojanz, I. Miccoli, C. Tegenkamp, M. Batzill, *ACS Nano* **2017**, *11*, 5130.

[200]　Y. Yu, G. Wang, S. Qin, N. Wu, Z. Wang, K. He, X.-A. Zhang, *Carbon* **2017**, *115*, 526.

[201]　W. Jolie, C. Murray, P. S. Weiß, J. Hall, F. Portner, N. Atodiresei, A. V. Krasheninnikov, C. Busse, H.-P. Komsa, A. Rosch, T. Michely, *Phys. Rev. X* **2019**, *9*, 011055.

[202]　M. Batzill, *J. Phys.: Condens. Matter* **2018**, *30*, 493001.

[203]　P. M. Coelho, H.-P. Komsa, H. Coy Diaz, Y. Ma, A. V. Krasheninnikov, M. Batzill, *ACS Nano* **2018**, *12*, 3975.

[204]　J. Dai, Z. Zhang, Z. Pan, C. Wang, C. Zhang, Z. Cheng, W. Ji, *Kagome bands and magnetism in MoTe$_{2-x}$ kagome monolayers*, arXiv, **2024**.

[205]　Z. Pan, W. Xiong, J. Dai, Y. Wang, T. Jian, X. Cui, J. Deng, X. Lin, Z. Cheng, Y. Bai, C. Zhu, D. Huo, G. Li, M. Feng, J. He, W. Ji, S. Yuan, F. Wu, C. Zhang, H.-J. Gao, *Ferromagnetism and correlated insulating states in monolayer Mo33Te56*, arXiv, **2024**.

[206]　Y.-H. Lin, C.-J. Chen, N. Kumar, T.-Y. Yeh, T.-H. Lin, S. Blügel, G. Bihlmayer, P.-J. Hsu, *Nano Lett.* **2022**, *22*, 8475.

[207]　Z.-M. Zhang, B.-C. Gong, J.-H. Nie, F. Meng, Q. Zhang, L. Gu, K. Liu, Z.-Y. Lu, Y.-S. Fu, W. Zhang, *Nano Lett.* **2023**, *23*, 954.

[208]　Q. Wu, W. Quan, S. Pan, J. Hu, Z. Zhang, J. Wang, F. Zheng, Y. Zhang, *Nano Lett.* **2024**, acs.nanolett.4c01526.

[209]　S. N. Magonov, P. Zoennchen, H. Rotter, H. J. Cantow, G. Thiele, J. Ren, M. H. Whangbo, *J. Am. Chem. Soc.* **1993**, *115*, 2495.

[210]　H. Zhang, Z. Shi, Z. Jiang, M. Yang, J. Zhang, Z. Meng, T. Hu, F. Liu, L. Cheng, Y. Xie, J. Zhuang, H. Feng, W. Hao, D. Shen, Y. Du, *Advanced Materials* **2023**, *35*, 2301790.

[211]　Z. Cai, H. Cao, H. Sheng, X. Hu, Z. Sun, Q. Zhao, J. Gao, S. Ideta, K. Shimada, J. Huang, P. Cheng, L. Chen, Y. Yao, S. Meng, K. Wu, Z. Wang, B. Feng, *Nano Lett.* **2024**, acs.nanolett.4c02580.

[212]　X. Xu, X. Wang, S. Yu, C. Wang, G. Liu, H. Li, J. Yang, J. Li, T. Sun, X. Hai, L. Li, X. Liu, Y. Zhang, W. Zhang, Q. Zhang, K. Wang, N. Xu, Y. Ma, F. Ming, P. Cui, J. Lu, Z. Zhang, X. Xiao, *High-density single-atom electrocatalytic centers on*

*two-dimensional topological platinum tellurides with Te-vacancy superstructure*, arXiv, **2024**.

[213] R. Wang, C. Wang, R. Li, D. Guo, J. Dai, C. Zong, W. Zhang, W. Ji, *High-Throughput Discovery of Kagome Materials in Transition Metal Oxide Monolayers*, arXiv, **2024**.

[214] H. Zhang, Q. Liu, L. Deng, Y. Ma, S. Daneshmandi, C. Cen, C. Zhang, P. M. Voyles, X. Jiang, J. Zhao, C.-W. Chu, Z. Gai, L. Li, *Nano Lett.* **2024**, *24*, 122.

[215] S. Duan, J.-Y. You, Z. Cai, J. Gou, D. Li, Y. L. Huang, X. Yu, S. L. Teo, S. Sun, Y. Wang, M. Lin, C. Zhang, B. Feng, A. T. S. Wee, W. Chen, *Nat Commun* **2024**, *15*, 8940.

[216] B. Zhu, W. Huang, H. Lin, H. Feng, K. Palotás, J. Lv, Y. Ren, R. Ouyang, F. Yang, *J. Am. Chem. Soc.* **2024**, *146*, 15887.

[217] Q. Liu, N. Han, S. Zhang, J. Zhao, F. Yang, X. Bao, *Nano Res.* **2018**, *11*, 5957.

[218] W. Huang, Q. Liu, Z. Zhou, Y. Li, Y. Ling, Y. Wang, Y. Tu, B. Wang, X. Zhou, D. Deng, B. Yang, Y. Yang, Z. Liu, X. Bao, F. Yang, *Nat Commun* **2020**, *11*, 2312.

[219] S. Wang, Z. Zhan, X. Fan, Y. Li, P. A. Pantaleón, C. Ye, Z. He, L. Wei, L. Li, F. Guinea, S. Yuan, C. Zeng, *Phys. Rev. Lett.* **2024**, *133*, 066302.

[220] S. Xing, T. Zhao, J. Zhou, Z. Sun, *J. Phys. Chem. C* **2024**, *128*, 2618.

[221] H. Zhou, M. dos Santos Dias, Y. Zhang, W. Zhao, S. Lounis, *Nat Commun* **2024**, *15*, 4854.

[222] D. Lee, K.-H. Jin, F. Liu, H. W. Yeom, *Nano Lett.* **2022**, *22*, 7902.

[223] M. G. Scheer, B. Lian, *Phys. Rev. Lett.* **2023**, *131*, 266501.

[224] J. Jung, Y.-H. Kim, *Phys. Rev. B* **2022**, *105*, 085138.

[225] A. Olariu, P. Mendels, F. Bert, F. Duc, J. C. Trombe, M. A. de Vries, A. Harrison, *Phys. Rev. Lett.* **2008**, *100*, 087202.

[226] M. P. Shores, E. A. Nytko, B. M. Bartlett, D. G. Nocera, *J. Am. Chem. Soc.* **2005**, *127*, 13462.

[227] L. Balents, *Nature* **2010**, *464*, 199.

[228] C. Zhang, Z. Pan, W. Xiong, J. Dai, Y. Wang, T. Jian, X. Cui, J. Deng, X. Lin, Z. Cheng, Y. Bai, C. Zhu, D. Huo, G. Li, M. Feng, J. He, W. Ji, S. Yuan, F. Wu, H.-J. Gao, *Spin-polarized correlated insulator in monolayer MoTe2-x*, **2023**.